\documentclass[useAMS,usenatbib,twocolumn]{mn2e}
\pdfoutput=1
\usepackage{amssymb}
\usepackage{amsmath}
\usepackage{natbib}
\usepackage[pdftex]{graphicx}
\usepackage{subfigure}
\usepackage{booktabs}

\begin{document}

\title[Wave breaking in a solar-type star]
{Three-dimensional simulations of internal wave breaking and the fate
  of planets around solar-type stars}
      \author[A.J. Barker]{Adrian J. Barker\thanks{E-mail:
	  ajb268@cam.ac.uk} \\
	Department of Applied Mathematics and Theoretical
	Physics, University of Cambridge, Centre for Mathematical Sciences, \\ Wilberforce Road,
	Cambridge CB3 0WA, UK}
	
\date{Accepted 2011 February 3.  Received 2011 January 26; in original
  form 2010 November 3}

\pagerange{\pageref{firstpage}--\pageref{lastpage}} \pubyear{2011}

\maketitle

\label{firstpage}

\begin{abstract}
  We study the fate of internal gravity waves approaching the centre
  of an initially non-rotating solar-type star, by performing
  three-dimensional numerical simulations using a Boussinesq-type
  model. These waves are excited at the top of the radiation zone by the tidal forcing of a
  short-period planet on a circular, coplanar orbit. This extends
  previous work done in two dimensions by Barker \& Ogilvie. We first
  derive a linear wave solution, which is not exact in 3D; however,
  the reflection of ingoing waves from the centre is close to perfect
  for moderate amplitude waves. Waves with sufficient amplitude to
  cause isentropic overturning break, and deposit their angular
  momentum near the centre. This forms a critical layer, at which the
  angular velocity of the flow matches the orbital angular frequency
  of the planet. This
  efficiently absorbs ingoing waves, and spins up the star from the
  inside out, while the planet spirals into the star.

  We also perform numerical integrations to determine the linearised
  adiabatic tidal response throughout the star, in a wide range of
  solar-type stellar models with masses in the range $0.5 \leq
  m_{\star}/M_{\odot} \leq 1.1$, throughout their main sequence
  lifetimes. The aim is to study the influence of the launching region for
  these waves at the top of the radiation zone in more detail, and to 
  determine the accuracy of a
  semi-analytic approximation for the tidal torque on the star, that
  was derived under the assumption that all ingoing wave angular momentum is absorbed in
  a critical layer. 

  The main conclusions of this work are that this nonlinear mechanism of tidal dissipation could 
  provide an explanation for the survival of all short-period
  extrasolar planets observed around FGK stars, while it predicts the
  destruction of more massive planets. This work provides
  further support for the model outlined in a previous paper by Barker \&
  Ogilvie, and makes predictions that will be tested by
  ongoing observational studies, such as WASP and Kepler.
\end{abstract}


\section{Introduction}

Tidal interactions are important in determining the fate of
short-period extrasolar planets and the spins of their host stars. The
extent of the spin-orbit evolution that results from tides depends on
the dissipative properties of the star and planet. These are usually parametrized
by a dimensionless quality factor for each body, which is an inverse
measure of the dissipation. This is usually defined to be
proportional to the ratio of the maximum energy stored in a tidal oscillation to the energy dissipated
over one cycle (e.g.~\citealt{GoldSot1966}). 

In principal the modified quality factor\footnote{Related to the
  traditional $Q$ by $Q^{\prime} =
3Q/2k$, where $k$ is the second-order potential Love number
of the body.} $Q^{\prime}$ depends on tidal
frequency, the internal structure of the body, and the amplitude of
the tidal forcing. Unfortunately, the mechanisms of dissipation that
contribute to $Q^{\prime}$ are poorly understood. To simplify this
difficult problem, nearly all studies neglect any amplitude dependence of
the dissipation (except, for example \citealt{GoodmanLackner2009}). Such studies already exhibit a complicated
dependence of $Q^{\prime}$ on the tidal frequency
(e.g~\citealt{SavPap1997}; \citealt{Wu2005b}; \citealt{Gio2004},
  hereafter OL04; \citealt{Gio2007},
  hereafter OL07; \citealt{PapIv2010}), and the internal
structure of the body (e.g~OL07; \citealt{Barker2009}). In this paper
we extend the work of \cite{Barker2010} (hereafter BO10) to study an
important nonlinear dissipation mechanism in solar-type stars.

We can decompose the tidal response of a fluid body, which in this
paper we take to mean a solar-type star, into an
equilibrium and a dynamical tide, defined such that the total
displacement is the sum of these two displacements, i.e, that
$\boldsymbol{\xi} = \boldsymbol{\xi}^{d} + \boldsymbol{\xi}^{e}$. The
equilibrium tide is a quasi-hydrostatic bulge defined by
\begin{eqnarray}
\label{eqmtideeqns}
\xi^{e}_{r} = -\frac{\Psi}{g}, \;\;\; \mathrm{and} \;\;\; \nabla \cdot
\boldsymbol{\xi}^{e} = 0,  
\end{eqnarray}
in stratified regions (\citealt{Goldreich1989}), where $\Psi$ is the tidal
gravitational potential experienced by the body and $g$ is the
gravitational acceleration. The total
displacement is not well described by Eqs.~\ref{eqmtideeqns} in convective
regions (\citealt{GoodmanDickson1998}, hereafter GD98;
\citealt{Terquem1998}, hereafter T98; OLO4). This is because in a barotropic flow
(with adiabatic stratification) vorticity is conserved, so we must have $\nabla \times
\boldsymbol{\xi} = 0$, whereas $\nabla \times \boldsymbol{\xi}^{e} \ne
0$, in general. The presence of a
convection zone (hereafter CZ) thus
implies that a dynamical tide must exist. The dynamical tide
$\boldsymbol{\xi}^{d}$ is defined as the residual response that
results from the equilibrium tide not being the exact (linearised)
solution to the problem, when the tidal frequency is nonzero. 

The equations governing
the adiabatic equilibrium and dynamical tides in linear theory are 
\begin{eqnarray}
\label{eqmtideeqn}
 0 &=& -\frac{1}{\rho}\nabla \delta p^{e} + \frac{\delta
   \rho^{e}}{\rho^{2}}\nabla p - \nabla \Psi, \\
-\omega^{2} \boldsymbol{\xi}^{d} &=& -\frac{1}{\rho}\nabla \delta p^{d} + \frac{\delta \rho^{d}}{\rho^{2}}\nabla p + \underbrace{\omega^{2}
  \boldsymbol{\xi}^{e}}_{\mathrm{Forcing}},
\label{dyntideeqn}
\end{eqnarray}
from which it is clear that the dynamical tide is not forced directly
by the tidal potential, only by the inertial terms in the equation of
motion.

Dynamical tides in radiation zones of solar-type
stars take the form of internal (inertia-) gravity waves (IGWs), which
have frequencies below the buoyancy (or Brunt-V\"ais\"al\"a) frequency $N$. These have
previously been proposed to contribute to $Q^{\prime}$ for early-type
stars (e.g.~\citealt{Zahn1975}), where they are damped at the surface by
radiative diffusion. It has also been proposed that these waves could
synchronise the spin of the star with the orbit (in this case of a
close-binary perturber) from the outside in \citep{Goldreich1989}. In
BO10, we considered a nonlinear mechanism of tidal
dissipation in solar-type stars (with radiative cores), extending an idea
by GD98. The consequences of this mechanism are similar to
\cite{Goldreich1989}, except that the star is synchronised with the
orbit from the inside out.

A short-period planet excites IGWs at the top of the RZ of a
solar-type star, where $N$ increases from zero with distance into the
RZ from the CZ/RZ interface. There is thus a location at which $N\sim 1/P$, with $P$
being the planetary orbital period, at which IGWs (which have
frequencies less than $N$) are efficiently excited. These waves propagate downwards into the
radiation zone (hereafter RZ), until they
reach the centre of the star, where they are geometrically focused
and can become nonlinear. If their amplitudes are sufficiently large, convective
overturning occurs, and the wave breaks. This has consequences for the
tidal torque, and the stellar $Q^{\prime}$. In this paper we study this
mechanism, primarily using three-dimensional numerical simulations.

In BO10, we derived a Boussinesq-type system of
equations that is relevant for describing the dynamics of IGWs
approaching the centre of a solar-type star. We then performed
numerical simulations, solving these equations in two dimensional cylindrical geometry. 
In this paper we extend these simulations to three dimensions, in
spherical geometry,
and confirm that the most important results of BO10 
are not affected by this extension. We first derive a linear wave
solution, which represents the waves excited by planets on
short-period orbits as they approach the
central $\lesssim 5\%$ of the star, within which $N \propto r$ and $g
\propto r$. A weakly nonlinear analysis
confirms that this solution is not exact, though the reflection of
ingoing waves is close to perfect for moderate amplitude waves. We
present the numerical setup and the analysis of the simulation results.
We then discuss the launching
region at the top of the RZ, and the possible effects of
magnetic fields on the problem. Together with BO10, we provide a possible explanation for
the survival of all short-period extrasolar planets observed thus far, which will
be tested by ongoing observational studies, such as WASP and Kepler.

\section{Tidal potential}

The tidal potential experienced by a star hosting a short-period planet can be written in standard
spherical polar coordinates $(r,\theta,\phi)$ as a sum of
rigidly rotating spherical harmonics
\begin{eqnarray}
\mathrm{Re}\left[\Psi_{l,m} r^{l} Y_{l}^{m}(\theta,\phi)e^{-i\omega t} \right],
\end{eqnarray}
in a non-rotating (but non-inertial) reference frame centred on the
star, where $Y_{l}^{m}$ is a spherical harmonic (normalised such that
the integral of $|Y_{l}^{m}|^2$ over solid angles is unity) and $\Psi_{l,m}$ is an amplitude. Here $\omega$ is the frequency in that frame, related to the
tidal frequency by $\hat{\omega} = \omega - m \Omega$, where $\Omega$
is the spin angular frequency of the star. In this paper we consider the waves excited by planets on circular,
coplanar orbits, relegating any studies of the waves excited by eccentric and inclined
planets to future work. In this case the dominant component of the tidal
potential is quadrupolar ($l=2$), with $m=2$, and takes the form
\begin{eqnarray}
\label{tidpot}
\Psi(r,\theta,\phi,t) = -\sqrt{\frac{6\pi}{5}}\frac{Gm_{p}}{a^{3}}r^{2}Y_{2}^{2}(\theta,\phi - \frac{\hat{\omega}}{2}t),
\end{eqnarray}
where $m_{p}$ is the planetary mass and $a$ is its orbital semi-major axis.
Since most short-period planets orbit faster than their stars spin, as
a result of stellar magnetic braking, we take the star to be
(initially) non-rotating, so that $\hat{\omega} = \omega = 2n$, where $n$ is
the orbital angular frequency of the planet.

\section{Linear theory}
\label{lineartheory3D}

In this section we derive a linear wave solution, starting from the adiabatic
Boussinesq-type system (BO10)
\begin{eqnarray}
\label{MainEqs1}
&&D \mathbf{u} = -\nabla q + \mathbf{r}b, \\
%%%%%%%%%%%%%%%%%%%%%%%%%%
&&D b + C^{2} \mathbf{r}\cdot\mathbf{u}  = 0, \\
\label{MainEqs2}
%%%%%%%%%%%%%%%%%%%%%%%%%%
&&\nabla \cdot \mathbf{u} = 0, \\
\label{MainEqs3}
%%%%%%%%%%%%%%%%%%%%%%%%%%
&&D = \partial_{t} + \mathbf{u}\cdot \nabla,
\end{eqnarray}
where $\mathbf{u}$ is the fluid velocity, $b$ is a buoyancy variable
(proportional to the entropy perturbation) and $q$ is a modified
pressure variable. This system of equation was derived in 2D but is equally valid in
3D. We note that in this model $N=C r$, where $C$ is a
constant that measures the strength of the stable stratification at
the centre. This model is valid in the innermost $\lesssim 3 \%$ of the
star, which contains multiple wavelengths for the gravity waves excited by
short-period planets. We define $\Omega_{p} = \omega/m$, which is the pattern
speed of the forcing, and equals the orbital angular frequency of
the planet. We furthermore use the
non-dimensionalisation that the unit of length is $\Omega_{p}/C$ and
the unit of time is $1/\Omega_{p}$.
Linearising about hydrostatic equilibrium
in 3D spherical geometry, we seek solutions steady in the frame
rotating at the angular rate $\Omega_{p}$, proportional to $e^{im\xi}$, where
$\xi=\phi-\Omega_{p} t$ is
the azimuthal coordinate in this frame. This leads to the following equations:
\begin{eqnarray}
\label{3deqs1}
&& -imu_{r} = -\partial_{r}q + rb, \\
\label{3deqs2}
&& -imu_{\theta} = -\frac{1}{r}\partial_{\theta}q, \\
\label{3deqs3}
&& -imu_{\phi} = -\frac{im}{r\sin\theta}q, \\
\label{3deqs4}
&& -imb = -ru_{r}, \\
\label{3deqs5}
&& \frac{1}{r^{2}} \partial_{r}(r^{2}u_{r}) +
\frac{1}{r\sin\theta}\partial_{\theta}(\sin\theta u_{\theta}) + \frac{im}{r\sin\theta}u_{\phi} = 0.
\end{eqnarray}
To obtain the linear solution, we expand scalar quantities (i.e.~$q$
and $b$) in spherical harmonics $Y_{l}^{m}(\theta,\xi)$, since the problem is separable in both angular
coordinates. In Eqs.~\ref{3deqs1}--\ref{3deqs5} we have already
included the $\xi$-dependence of these functions ($e^{im\xi}$). We
thus take $q = \hat{q}(r)Y_{l}^{m}(\theta,0)$,
so that the total functions are expanded onto
$Y_{l}^{m}(\theta,\xi)$, and similarly for $b$ and $u_{r}$, where from now on we drop the hats on the
radial functions. The remaining velocity components are expanded onto
angular functions as appropriate to satisfy
Eqs.~\ref{3deqs1}--\ref{3deqs5}. The relation
\begin{eqnarray}
  q = \frac{im}{l(l+1)} \partial_{r}(r^{2}u_{r}),
\end{eqnarray}
follows from incompressibility and the result
\begin{eqnarray}
  \left[\frac{1}{\sin \theta} \partial_{\theta} \sin \theta \partial_{\theta}Y_{l}^{m} +
    \frac{1}{\sin^{2}\theta} \partial^{2}_{\phi}Y_{l}^{m}\right]= -l(l+1)Y_{l}^{m}.
\end{eqnarray}
This enables us to derive the linear differential equation
\begin{eqnarray}
\partial_{r}^{2} (r^{2} u_{r}) - \frac{l(l+1)}{m^{2}}(m^{2}-r^{2})u_{r} = 0,
\end{eqnarray}
whose solutions can be written in terms of Bessel functions of
half-integer order (alternatively these can be written as spherical
Bessel functions, or they can be reduced to elementary functions).
The corresponding (total) linear solution for a standing wave in 3D can be written (where
real parts are assumed to be taken)
\begin{eqnarray}
\label{linearsoln3D}
u_{r}(r,\theta,\xi) \hspace{-0.2cm} &=& \hspace{-0.2cm} Br^{-\frac{3}{2}}J_{l+\frac{1}{2}}\left(k r\right)Y_{l}^{m}(\theta,\xi), \\
u_{\theta}(r,\theta,\xi) \hspace{-0.2cm} &=& \hspace{-0.2cm} \frac{B}{l(l+1)}
\frac{1}{r}\partial_{r}\left[r^{\frac{1}{2}}J_{l+\frac{1}{2}}\left(k r\right)\right]
\partial_{\theta} Y_{l}^{m}(\theta,\xi), \\
u_{\phi}(r,\theta,\xi) \hspace{-0.2cm} &=& \hspace{-0.2cm} \frac{imB}{l(l+1)}\frac{1}{r}\partial_{r}
\left[r^{\frac{1}{2}}J_{l+\frac{1}{2}}\left(k r\right)\right]
\frac{1}{\sin \theta} Y_{l}^{m}(\theta,\xi), \\
b(r,\theta,\xi) \hspace{-0.2cm} &=& \hspace{-0.2cm}
-\frac{iB}{m}r^{-\frac{1}{2}}J_{l+\frac{1}{2}}\left(k
  r\right)Y_{l}^{m}(\theta,\xi),
\label{linearsoln3D1}
\end{eqnarray}
where $B \in \mathbb{C}$ is an amplitude, and
\begin{eqnarray}
  k = \frac{1}{m}\sqrt{l(l+1)}.
\end{eqnarray}
Ingoing wave (hereafter IW) and outgoing wave (hereafter OW) solutions can be obtained by replacing the
Bessel function of the first kind by equivalent Hankel functions of
the first ($J_{\nu}+iY_{\nu}$) and second kinds ($J_{\nu}-iY_{\nu}$), respectively. 
Note also that
\begin{eqnarray}
\nonumber
 && \hspace{-0.5cm} \frac{1}{r}\partial_{r}\left[r^{\frac{1}{2}}\left(J_{l+\frac{1}{2}}\left(k r\right) \pm 
  iY_{l+\frac{1}{2}}\left(k r\right)\right)\right] = \\ &&
\hspace{0.5cm} \nonumber
r^{-\frac{3}{2}}\left[(1+l)\left(J_{l+\frac{1}{2}}\left(k r\right) \pm  
  iY_{l+\frac{1}{2}}\left(k r\right)\right)\right. \\ && \left. \hspace{0.5cm}
- kr\left(J_{(l+1)+\frac{1}{2}}\left(k r\right) \pm 
  iY_{(l+1)+\frac{1}{2}}\left(k r\right)\right)\right].
\end{eqnarray}

Starting from Eq.~46 of BO10, we can calculate a conserved
energy flux. Integrating this equation over $\xi$ eliminates the terms
containing derivatives in $t$ and $\phi$ due to periodicity in $\xi$,
as a result of the fundamental theorem of calculus. This
allows the definition of a conserved quantity proportional to the
energy flux,
\begin{eqnarray}
F = \int_{0}^{\pi} \int_{0}^{2\pi} r^{2}\sin\theta
  u_{r}(E+q)d\xi d\theta,
\end{eqnarray}
where the energy density per unit volume is $E =
\frac{1}{2}\rho_{0}\left(|\mathbf{u}|^{2} + \frac{b^{2}}{C^{2}}\right)$, with
  $\rho_{0}$ the (constant) central density of the star.
For linear waves, terms involving products are small, so
$E\ll q$, to this order. This leaves
\begin{eqnarray}
\label{flux1}
F = \pi r^{2} \int_{0}^{\pi} \mathrm{Re}\left[ u_{r} q^{*} \right]
\sin \theta d\theta.
\end{eqnarray}
Whether this is positive or negative depends on whether the wave is ingoing or outgoing.

Substituting the linear solution into
Eq.~\ref{flux1} provides a simple expression for the flux of a single
$l,m$ wave:
\begin{eqnarray}
\label{fluxsimple}
F = \frac{m}{\pi l(l+1)}(|A_{out}|^{2}-|A_{in}|^{2}), 
\end{eqnarray}
with corresponding energy flux $F^{E} = \rho_{0}F$, and angular
momentum flux $F^{L} = \frac{\rho_{0}m}{\omega}F$. We define
  the complex amplitudes of the IW and OW to be $A_{in}$ and
  $A_{out}$, respectively.
Eq.~\ref{fluxsimple} follows from the orthonormality of spherical harmonics and the
Wronskian of the Hankel functions
\begin{eqnarray}
&& \int_{0}^{2\pi}\int_{0}^{\pi}Y_{l}^{m}(\theta,\xi) [Y_{l^{\prime}}^{m^{\prime}}(\theta,\xi)]^{*} \sin
\theta d\theta d\xi= \delta_{l}^{l^{\prime}}\delta_{m}^{m^{\prime}}, \\
&& W[J_{\nu}(kr)+iY_{\nu}(kr),J_{\nu}(kr)-iY_{\nu}(kr)]=\frac{4}{i\pi
  r}.
\end{eqnarray}
The particular wave that we will study has $l=m=2$, and the components 
\begin{eqnarray}
\label{lm2wave}
u_{r}(r,\theta,\xi) &=& Br^{-\frac{3}{2}}J_{\frac{5}{2}}\left(k r\right)Y_{2}^{2}(\theta,\xi), \\
u_{\theta}(r,\theta,\xi) &=& \frac{B}{l(l+1)}
\frac{1}{r}\partial_{r}\left[r^{\frac{1}{2}}J_{\frac{5}{2}}\left(k r\right)\right]
\partial_{\theta} Y_{2}^{2}(\theta,\xi), \\
u_{\phi}(r,\theta,\xi) &=& \frac{imB}{l(l+1)}\frac{1}{r}\partial_{r}
\left[r^{\frac{1}{2}}J_{\frac{5}{2}}\left(k r\right)\right]
\frac{1}{\sin \theta} Y_{2}^{2}(\theta,\xi), \\
b(r,\theta,\xi) &=& -\frac{iB}{m}r^{-\frac{1}{2}}J_{\frac{5}{2}}\left(k r\right)Y_{2}^{2}(\theta,\xi),
\end{eqnarray} 
where $k=\sqrt{6}/2$.

\subsection{Criterion for overturning isentropes}

A condition in 3D for isentropic overturning can be derived from
considering when the radial gradient of the entropy $s = b +
(1/2)r^{2}$ becomes negative, i.e., when
  $(1/r)\partial_{r} b < -1$. This is equivalent to the
  criterion that $|(1/r)\partial_{r} (r u_{r})| > |m|$, which can be
  shown by using Eq.~\ref{3deqs4}. 
To correlate our notation with the appendix of OL07, we define the dimensionless nonlinearity
parameter $A$, such that overturning occurs if $|A|>1$. This is
defined such that the radial velocity is the real part of
\begin{eqnarray}
\nonumber
&& \hspace{-0.5cm} u_{r} = 40Ar^{-4}\left[\frac{1}{\sqrt{6}}\left(1-\frac{1}{2}r^{2}\right)\sin
  k r - \right. \\ && \hspace{4cm} \left. \frac{1}{2} r \cos k r\right]\sin^{2}\theta e^{i m \xi},
\end{eqnarray}
in the dimensionless units that we have been using in this section
(this is equivalent to Eq.~\ref{lm2wave}).

Overturning is achieved at the centre
when the radial velocity in the wave takes a maximum value $u_{r} = 1.27
$ at its innermost peak at $r=2.04$. This value is used to compare
it to the magnitude of the radial velocity achieved in numerical simulations at the onset of wave
breaking. In these dimensionless units, we similarly require
$u_{\phi} \gtrsim 0.99$ or $b \gtrsim 0.38$. Note that there
is not such a simple interpretation of the criterion on $u_{\phi}$ as
in 2D, though the value is quantitatively very similar\footnote{In the
  figures below, we use a different nondimensionalisation of the
  velocity, by normalising it to the constant (asymptotic) radial
  phase velocity $\omega\lambda_{r}/2\pi$ in the wave. this is
  identical to BO10, and gives values exactly 1/2 of those in the
  units of this section.}. From the results of
the 2D simulations in BO10, we expect the waves to undergo instability and
break within several wave periods after these criteria begin to be satisfied.

\subsection[Weakly nonlinear theory]{Weakly nonlinear theory\footnote{I would like to thank
    Gordon Ogilvie for suggesting the calculations in this section.}}
\label{weaklynonlinear}

The linear solution written down in Eqs.~\ref{linearsoln3D}--\ref{linearsoln3D1} is not a nonlinear
solution, unlike the equivalent in 2D. This can be shown by computing the nonlinear terms in the
full Boussinesq-type system using the linear solution, in
Mathematica, for example. We find that $\mathbf{u} \cdot \nabla b
\ne0$, in general, and similarly for the nonlinear terms in
the momentum equation. This means that the reflection of the waves from the centre could be
different than in 2D, since nonlinearities do not vanish for this
wave. In this section we perform a weakly nonlinear analysis to determine the
dominant nonlinear effects for small amplitudes. Since these
nonlinearities do not vanish, this highlights the importance of numerical simulations for
these waves approaching the centre. We describe the results of such
simulations in \S~\ref{3dresults}.
 
We propose a weakly nonlinear solution of the form
\begin{eqnarray}
\nonumber
u_{r}(r,\theta,\xi) &=& \frac{\epsilon}{2} \{u_{r1}(r,\theta)e^{i\xi} +
u_{r1}^{*}(r,\theta)e^{-i\xi}\} \\
\nonumber
&&+\epsilon^{2}\{ u_{r20}(r,\theta) \\ && \nonumber +
\frac{1}{2}\left(u_{r22}(r,\theta)e^{2i\xi} +
  u_{r22}^{*}(r,\theta)e^{-2i\xi}\right)\} 
\\ && + O(\epsilon^{3}),
\end{eqnarray}
and similarly for the other variables, where $\epsilon \ll
1$. This form is adopted because we are interested in
  calculating whether the incoming wave generates harmonics through
  the quadratic (self-)nonlinearities. These
additional waves (other than $m=0$) will escape to infinity and carry away a
portion of the energy flux. Here we write
$u_{r1}(r,\theta)e^{i\xi} = u_{r}(r,\theta,\xi)$ from Eq.~\ref{lm2wave}, for the $l=m=2$ wave
above, and similarly for other variables. We substitute these expansions into
the Boussinesq-type system and equate powers of $\epsilon$. 
At each order we also equate coefficients of $e^{in\xi}$. At leading
order only one mode is present, and we obtain the previously derived linear solution. 
After some algebra the solution at
$O(\epsilon^{2})$ can be computed to give
\begin{eqnarray}
\nonumber
u_{r22}(r,\theta,\xi) &=&
  A_{22}r^{-\frac{3}{2}}\left[J_{9/2}\left(\sqrt{5/6}k r\right)
  \right. \\ && \left. +iY_{9/2}\left(\sqrt{5/6}k
  r\right)\right]Y_{4}^{4}(\theta,\xi),
\label{leq4meq4wave}
\end{eqnarray}
which is an $l=m=4$ wave with complex amplitude $A_{22}$, which
has been computed using Mathematica and given in terms of $A$.

For the wave described by Eq.~\ref{leq4meq4wave}, $F$ can be
computed. The ratio of the energy flux in the outgoing $l=m=4$ wave
to the ingoing $l=m=2$ wave can be shown to be approximately $1.2
\times 10^{-5}|A|^{2}$. We define a reflection coefficient
\begin{eqnarray} 
\mathcal{R} = \left| \frac{A_{out}}{A_{in}} \right|,
\end{eqnarray}
which measures the amplitude decay for a wave travelling from a radius
$r$ to the centre, and back to $r$. For perfect reflection from the centre
$\mathcal{R} = 1$, whereas complete absorption means that $\mathcal{R} = 0$. The reflection coefficient for reflection from the centre, for a weakly nonlinear
$l=m=2$ wave, can be computed from
\begin{eqnarray} 
\mathcal{R}^{2} \approx 1 - 1.2 \times 10^{-5}|A|^{2}. 
\end{eqnarray}
This means that a weakly nonlinear primary wave (with $|A| \ll
1$), will reflect approximately perfectly from the centre, with a reflection coefficient
that is close to unity. However, a small fraction of the IW energy
flux is transferred to waves with higher $l$ and $m$-values,
reinforcing the fact that Eqs.~\ref{linearsoln3D}--\ref{linearsoln3D1} is not an exact solution,
contrary to the analogous solution in 2D.

\section{Numerical setup}

We solve the Boussinesq-type system in three dimensions using the
Cartesian pseudospectral code SNOOPY \citep{Lesur2005}, as in
BO10 (see \S~6 of that paper for further details). However, we modify the forcing and damping to take into account the
$z$-direction, and instead of forcing an $m=2$ wave in the equatorial
plane, we now have 
\begin{eqnarray}
\nonumber
\mathbf{f} \hspace{-0.1cm} &=& \hspace{-0.1cm} - f_{r}\mathrm{Re}\left[Y_{2}^{2}(\theta,\phi -
  \frac{\omega}{2}  t)\right]\mathbf{e}_{r}, \\ \hspace{-0.1cm} &=& \hspace{-0.1cm} - f_{r}
\frac{1}{4}\sqrt{\frac{15}{2\pi}}\frac{1}{r^{2}}\left\{
  \left(x^{2}-y^{2}\right)\cos \omega t - 2xy\sin \omega t \right\} \mathbf{e}_{r},
\end{eqnarray}
in Cartesian coordinates (with $r^{2} = x^{2}+y^{2}+z^{2}$), which is
applied in the region $0.85r_{box} \leq r \leq 0.9r_{box}$. We study a region $x,y,z \in
[-r_{box},r_{box}]$, where $r_{box}=1.5$, in arbitrary units (not the
same as \S~\ref{lineartheory3D}). For $r > 0.9r_{box}$ the
solution is damped to zero by using a parabolic smoothing function, as
in BO10.

We primarily use a resolution of $256^{3}$, for which the simulations were possible to run
on a single Intel Core i7 machine, utilising all 8 cores, with a
typical run time of several weeks to
resolve a hundred wave crossing times. We set $\omega = 1$ and choose a
typical IGW wavelength of $\lambda_{r} = 0.15$, giving approximately 8
wavelengths within the box. The value of $\lambda_{r}$ is
chosen to be slightly larger than that used in most of the 2D
calculations. This increases the number of grid points within a
wavelenth of the primary wave, partially offsetting the reduction in
resolution that results from using a smaller number of grid points per
dimension than in 2D. Choosing larger wavelengths than this is found
to result in unwanted effects from the proximity of the forcing
region, which modifies the linear solution. The viscosity and
radiative diffusion coefficients, at least one of which must be implemented in the code
for numerical stability, are chosen to be $ \nu = 4 \times
10^{-6}$ and $\kappa = 0$. An otherwise identical setup is used as in
BO10, except using spherical geometry instead of
cylindrical geometry.  We have confirmed that the
linear solution is well reproduced with this numerical setup.

\section{Numerical results}
\label{3dresults}

In this section we describe the results of the numerical
simulations. We analyse the results of the simulations using the IW/OW
decomposition described in Appendix \ref{refcoeff}, and reconstruct
the solutions from the computed IW/OW amplitudes to compare with the
simulation data, as described therein. From preliminary
investigation, we find that $\tilde{f}_{r}
\equiv f_{r} 2\pi/(\omega^{2}\lambda_{r}) \gtrsim 0.2$ is required for
breaking, so a variety of simulations are performed with forcing
amplitudes either side of this value. The basic results of these simulations are that the wave reflects
approximately perfectly from the centre of the star if the amplitude
of the wave is smaller than a certain critical value, which is found
to correspond with that required for isentropic overturning. Above
this value, wave breaking and critical layer formation occur. This
picture is identical to that in 2D.

\subsection{Low-amplitude simulations}

\begin{figure*}
  \begin{center}
  \hspace*{-1.5cm}
    \subfigure{
      \includegraphics[width=1.1\textwidth]{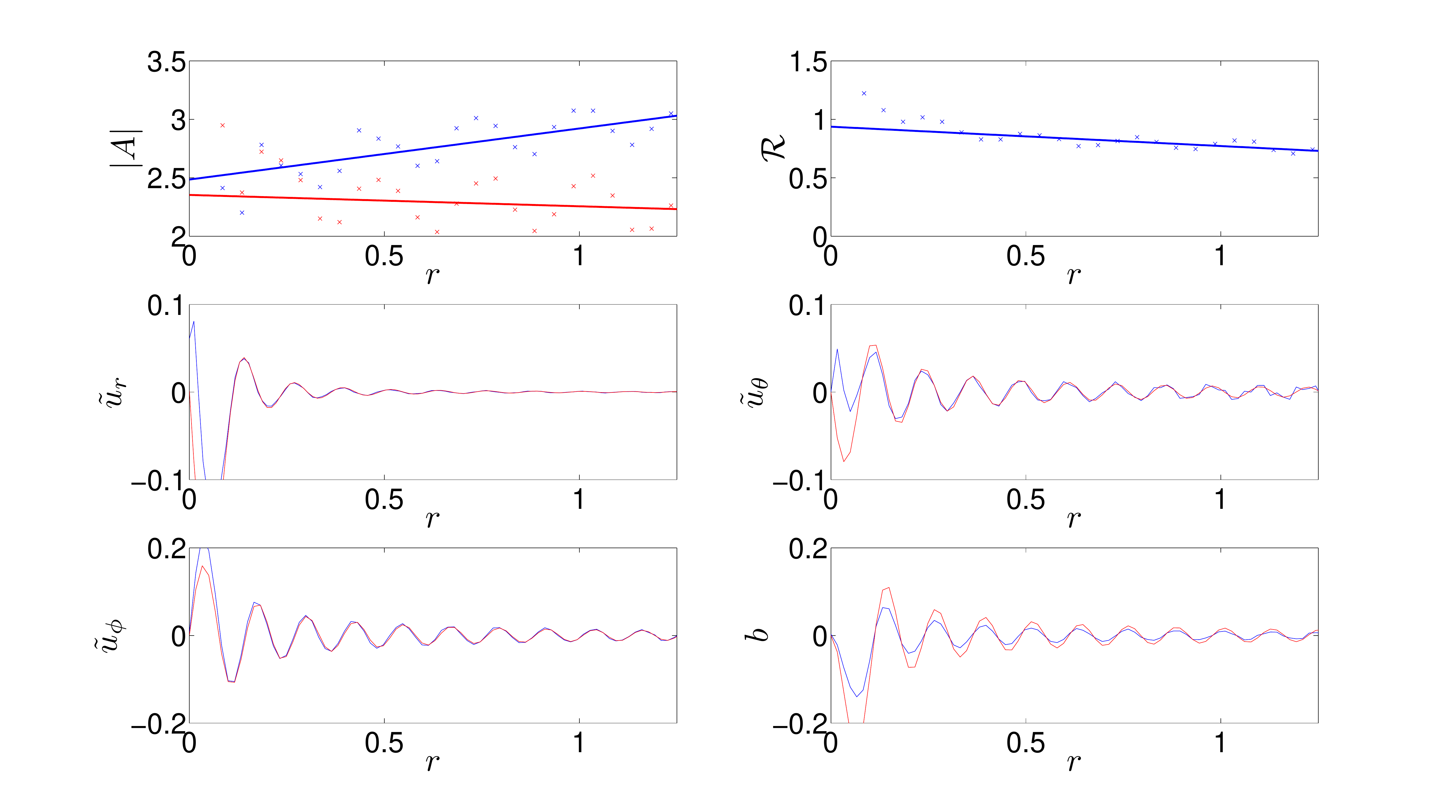} 
    }
  \end{center}
  \caption{In the top left panel we plot $A_{in}$ (blue) and $A_{out}$
    (red) vs. $r$, from a low-amplitude
      simulation in which the primary wave approximately perfectly reflects from
      the centre, with $\mathcal{R}$ displayed in the top right
      panel. Below, are the velocity components and the buoyancy variable (blue),
      plotted together with the reconstructed linear solution (red), against
      radius. This is taken from a simulation with $\tilde{f}_{r}=0.1$
      at $t=250$, once standing waves have formed. The amplitude decay
      with radius can be explained as due to viscous damping.
  }
\label{256lam15fr01t49AinAout}
\end{figure*}

\begin{figure}
  \begin{center}
    \subfigure{% trim=l b r t
      \includegraphics[width=0.48\textwidth,clip=true,trim=90mm 1mm 90mm 1mm]{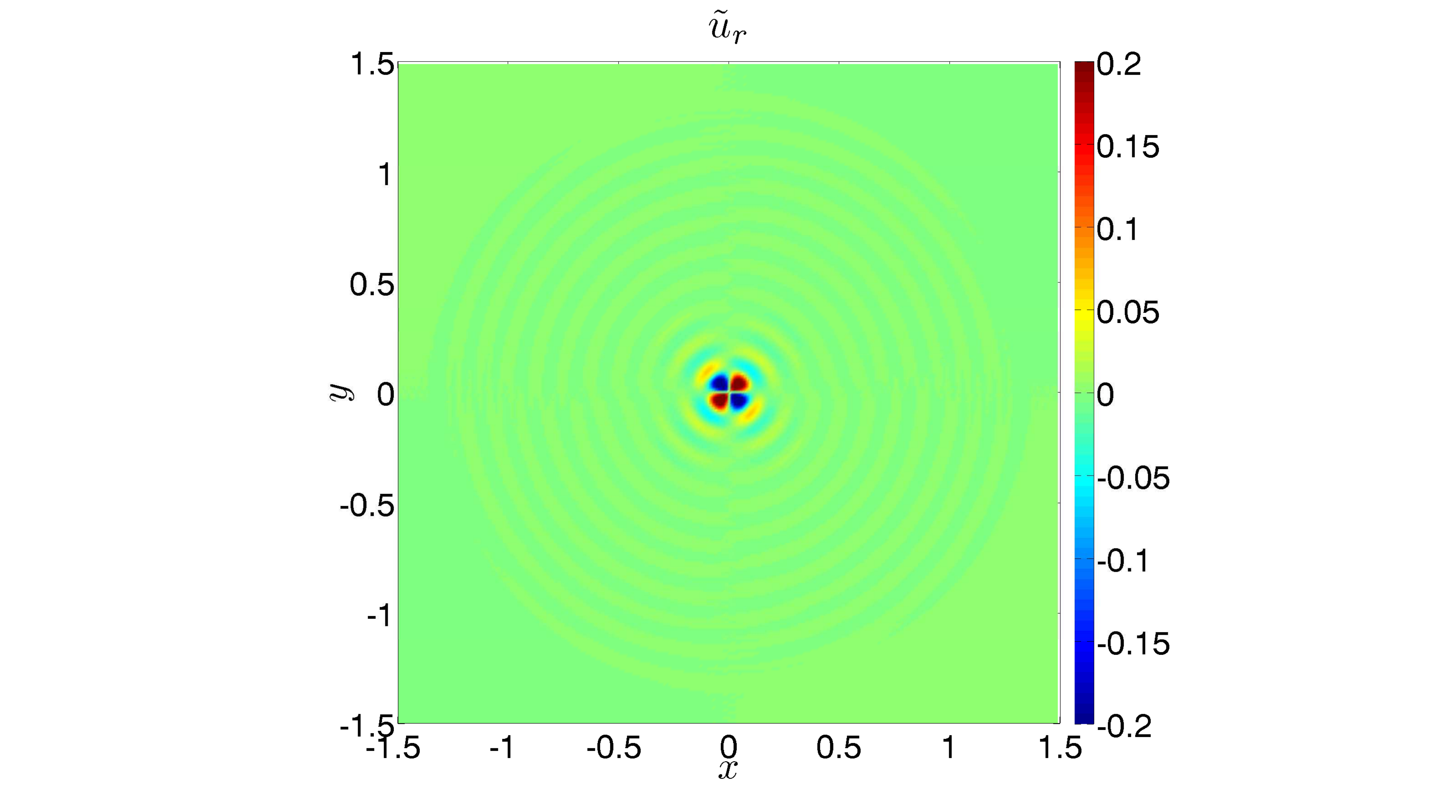} 
    }
    \subfigure{
      \includegraphics[width=0.48\textwidth,clip=true,trim=90mm 1mm 90mm 1mm]{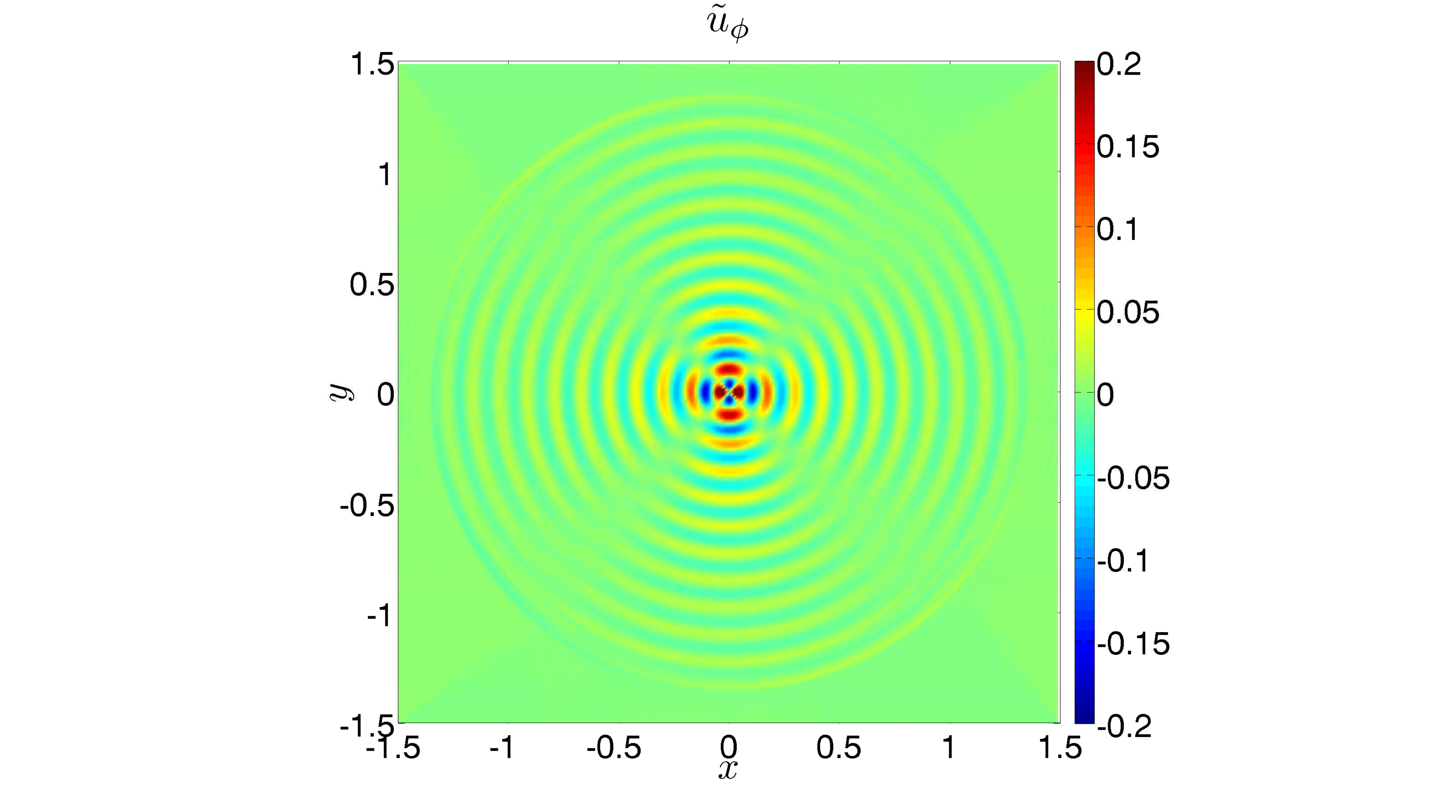} 
    }
  \end{center}
  \caption{2D plots of $\tilde{u}_{r}$ (top) and $\tilde{u}_{\phi}$
    (bottom) on the $xy$-plane for a simulation in which the wave perfectly reflects
    from the centre, for which $\tilde{f}_{r} = 0.1$ This can be
    qualitatively compared with Fig.~3 in BO10.
  }
\label{256lam15fr01t35xy}
\end{figure}

\begin{figure}
  \begin{center}
    \subfigure{
      \includegraphics[width=0.475\textwidth,clip=true,trim=90mm 1mm 90mm 1mm]{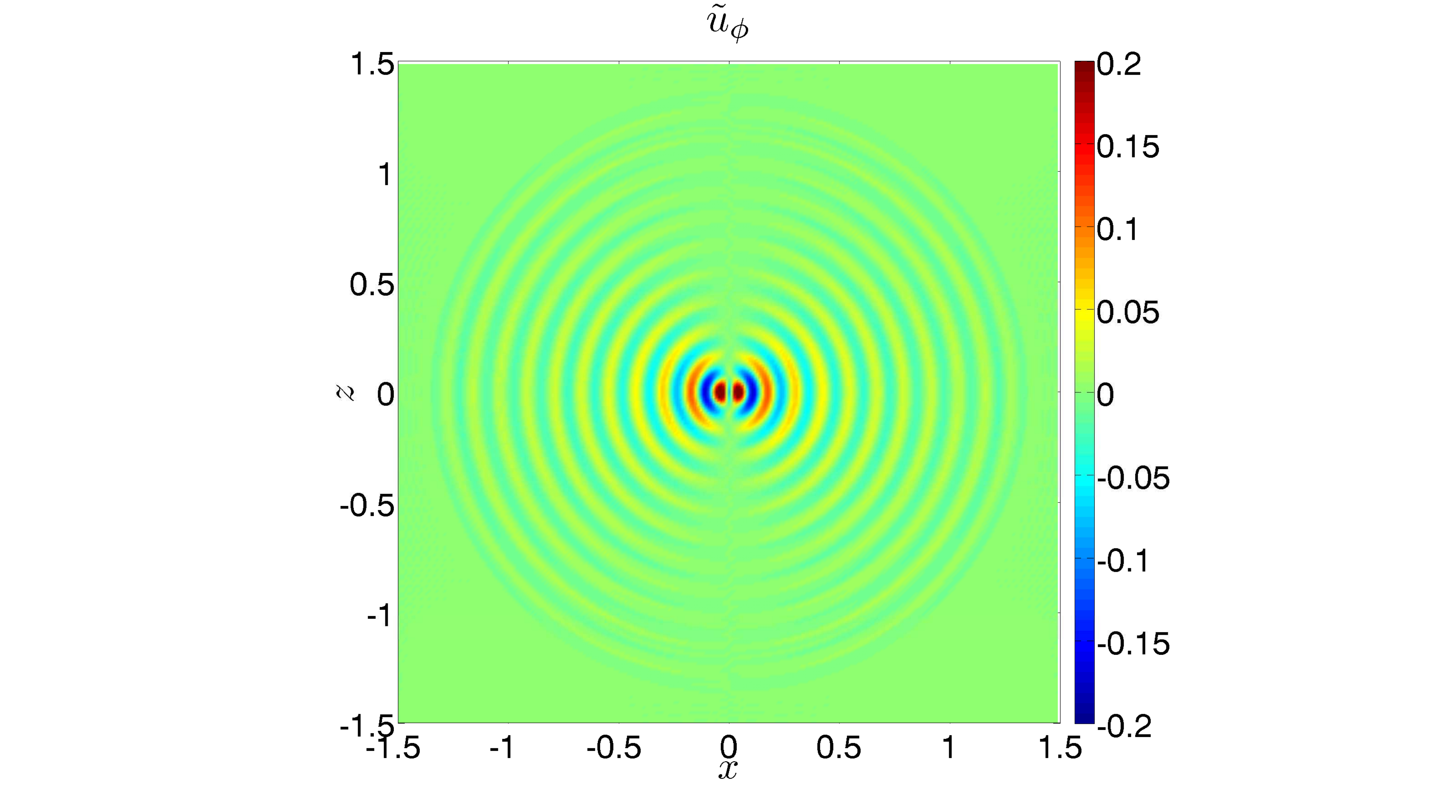} 
    }
  \end{center}
  \caption{2D plot of $\tilde{u}_{\phi}$ on the $xz$ plane
    for a simulation in which the wave perfectly reflects
    from the centre, for which $\tilde{f}_{r} =
    0.1$. $\tilde{u}_{\phi}$ is of largest magnitude
    in the equatorial plane, where $|Y_{2}^{2}|$ peaks.
  }
\label{256lam15fr01t35xz}
\end{figure}

For a low-amplitude simulation, with max($\tilde{u}_{r}$) below the
critical value for isentropic overturning (using $\tilde{f}_{r} =
0.1$), we plot the variation in amplitudes of the IWs and OWs, and
also the reconstructed solutions in Fig~\ref{256lam15fr01t49AinAout}.
We analyse the results using the method described in Appendix
\ref{refcoeff}, choosing a time once transients have been sufficiently
damped and standing waves have formed. The decay in radius is roughly
(though slightly smaller than) that which would be expected from
viscous damping, by the fractional amount
\begin{eqnarray}
\label{viscousdamping}
\frac{u_{r}}{u_{r,0}} \approx \exp \left(-2\int_{0}^{r} \frac{\nu
  k^{2}}{c_{g,r}} dr\right),
\end{eqnarray}
where $c_{g,r} = C\lambda_{r}^{2}/(2\pi^{2})$ is the (constant
asymptotic) radial group velocity, for a wave of the given wavelength.  As
in 2D, we have confirmed this explanation by running simulations
without viscosity, which are found to not exhibit this decay (though
these simulations eventually become numerically unstable if $\nu=\kappa=0$). Using a
smaller viscosity is also found to reduce the wavelength-scale
oscillations around the mean slope. These result from the fact that
the linear solution to the forced wave problem is no longer exact in
the presence of viscosity.  We find that increasing the number of grid
points within each shell (by reducing the values of
$i_{step},j_{step}$ and $k_{step}$) slightly reduces the vertical
extent of these oscillations, because this averages out the errors
that result from the assumption that the inviscid linear solution is
exact. However, increasing the number of grid points within
  each shell has negligible effect on the mean slope.

The spatial structure of the solutions in three dimensions in the $xy$-plane is very similar to that in two
dimensions, as can be seen in Fig.~\ref{256lam15fr01t35xy} (which can
be compared with Fig.~3 in BO10). In Fig.~\ref{256lam15fr01t35xz}, we
plot $\tilde{u}_{\phi}$ on the $xz$-plane. This shows that the magnitude
of the azimuthal velocity peaks at $\theta = \pi/2$, due to the latitudinal form of
$Y_{2}^{2}$.

From the calculation in \S~\ref{weaklynonlinear}, we expect the effects of nonlinearity to
be much weaker than the effects of viscosity for small-amplitude waves which do not
cause convective overturning. Since the effects of
weak nonlinearity are very small, it is difficult to
quantitatively confirm the results in
\S~\ref{weaklynonlinear}, using, for example, an extension of the
method described in Appendix \ref{refcoeff} for multiple $l$ and $m$ values. Nevertheless, we have qualitatively
confirmed the result that the reflection is coherent and nearly
perfect (in that $\mathcal{R} \approx 1$) for
amplitudes below that required for overturning the stratification. As
in 2D we do not observe any instabilities that act on the waves when
they have insufficient amplitude to overturn the stratification. In
this case, the waves can form global modes in the RZ.

\subsection{High-amplitude simulations}

\begin{figure*}
  \begin{center}
    \hspace*{-1.5cm}
    \subfigure{
      \includegraphics[width=1.1\textwidth]{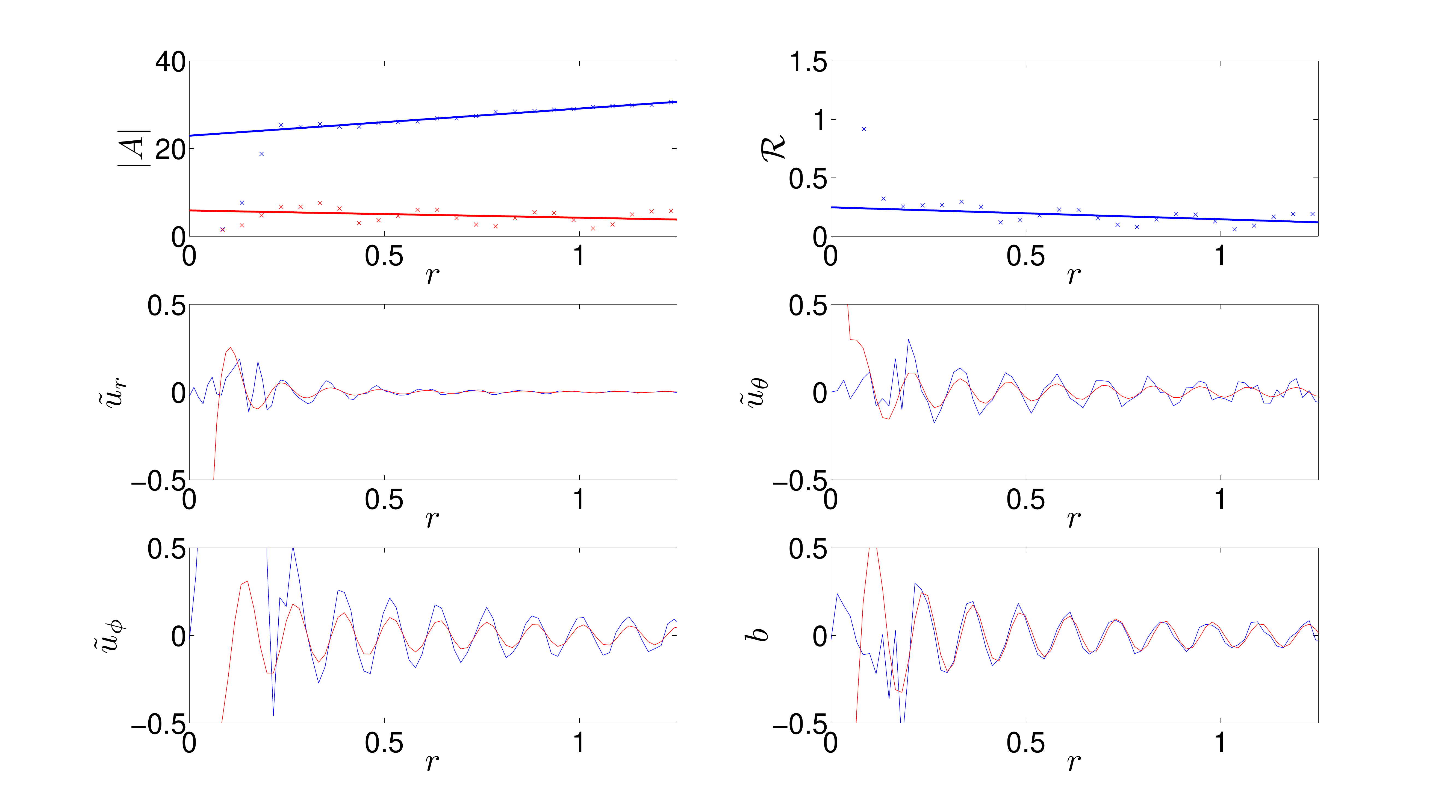} 
    }
  \end{center}
  \caption{In the top left panel we plot $A_{in}$ and $A_{out}$ vs. $r$ in a high-amplitude
      simulation in which the primary wave breaks ($\tilde{f}_{r}=1$)
      at $t=450$, with $\mathcal{R}$ in the top right panel. Below
      these are the velocity components and the buoyancy variable (blue),
      plotted together with the reconstructed linear solution (red), against
      radius. The central regions are not well described by the linear
      model, as expected.
  }
\label{256lam15fr1t45AinAout}
\end{figure*}

In high-amplitude simulations, in which the wave amplitude exceeds the
overturning criterion, the wave overturns the stratification during
part of its cycle and a rapid instability acts on the wave, which leads to
wave breaking within $1-3$ wave periods. This causes the rapid
(within several wave periods) deposition of primary wave angular
momentum, which spins up the mean flow to $\Omega_{p}$ (which
corresponds with the
orbital angular frequency of the planet) and produces
a critical layer. This critical layer acts as an
absorbing barrier for IWs, as is shown from
Fig.~\ref{256lam15fr1t45AinAout}, which plots the variation in
amplitude of the IW and OW, and also the reconstructed
wave solutions (which can be contrasted with
Fig.~\ref{256lam15fr01t49AinAout}). Once the critical layer has
formed, we find $|A_{out}| \ll |A_{in}|$. The central regions are
not well described by the linear model, as we
would expect. However, the region outside of $r\approx 0.25$ is well
described by the linear solution,  with $|A_{out}| \ll |A_{in}|$. In
this region, $\mathcal{R} \ll 1$, so it is reasonable to assume that
the IWs are efficiently absorbed near the centre. This picture is identical to that in 2D.

\begin{figure}
  \begin{center}
    \subfigure{% trim=l b r t
      \includegraphics[width=0.48\textwidth,clip=true,trim=90mm 1mm 90mm 1mm]{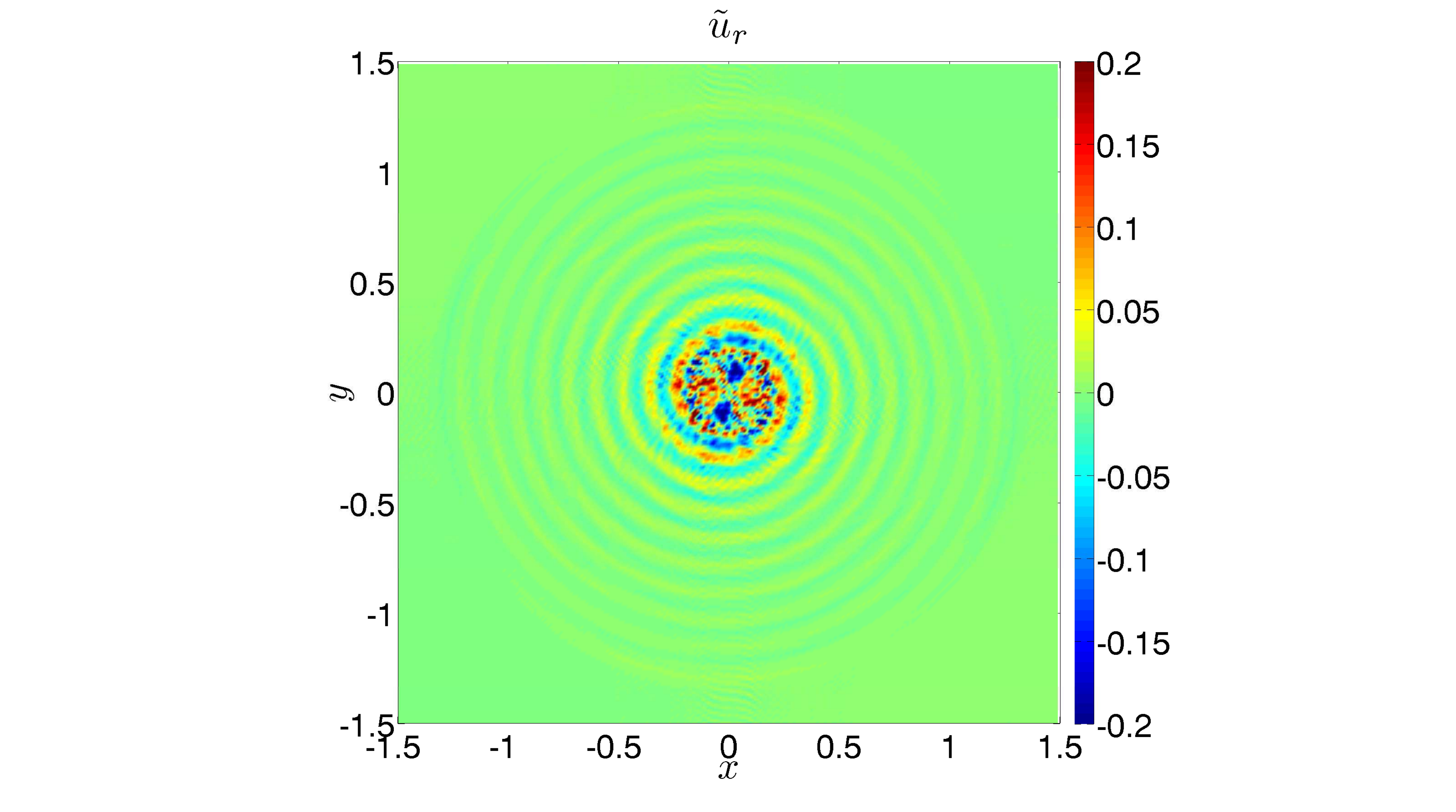} 
    }
    \subfigure{
      \includegraphics[width=0.48\textwidth,clip=true,trim=90mm 1mm 90mm 1mm ]{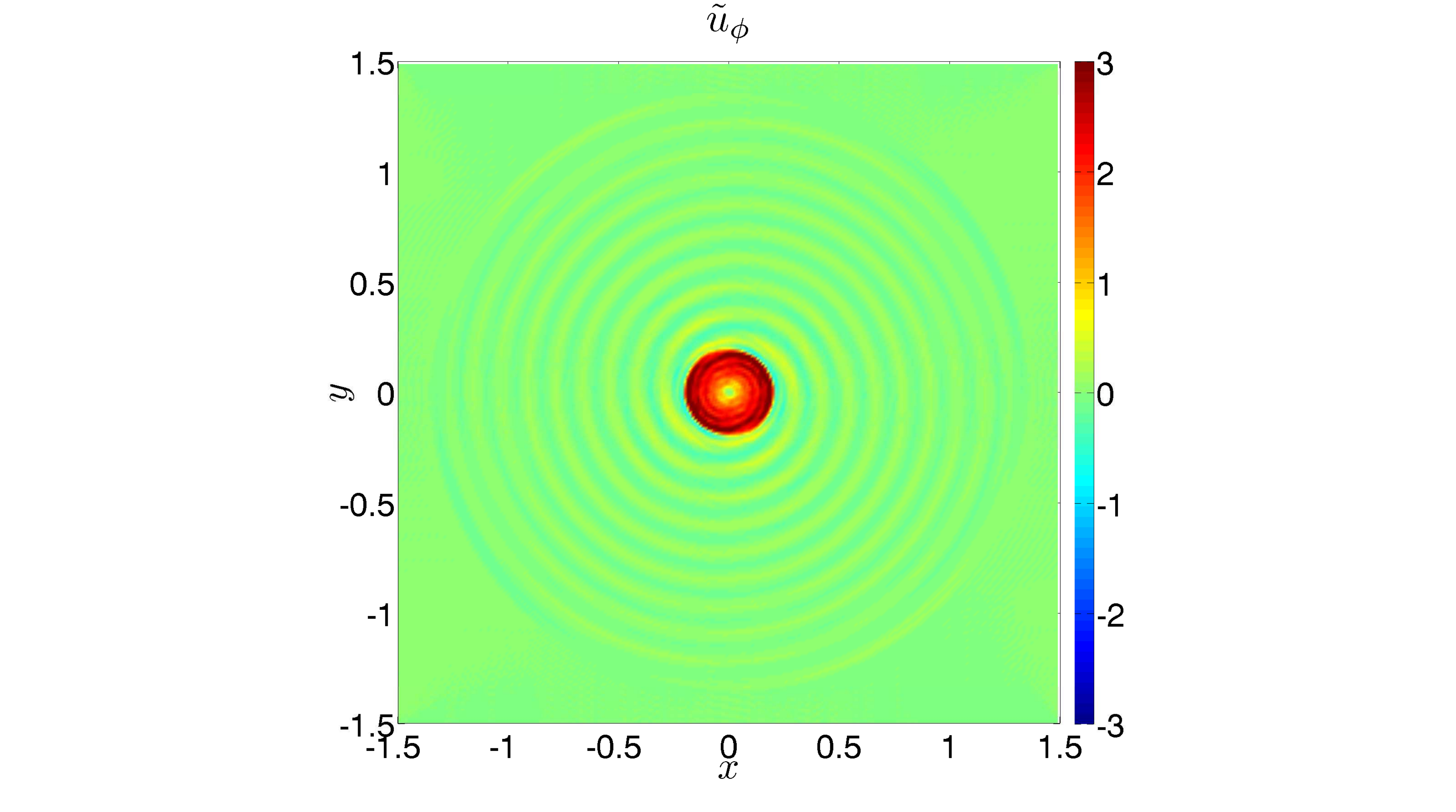} 
    } 
  \end{center}
  \caption{2D plot of $\tilde{u}_{r}$ (top) and $u_{\phi}$ (bottom) on $xy$-plane for a simulation in
    which breaking occurs with $\tilde{f}_{r}=1$, at $t=450$.
  }
  \label{256lam15fr1t45xy1}
\end{figure}

\begin{figure}
  \begin{center}
    \subfigure{
      \includegraphics[width=0.475\textwidth,clip=true,trim=90mm 1mm 90mm 1mm]{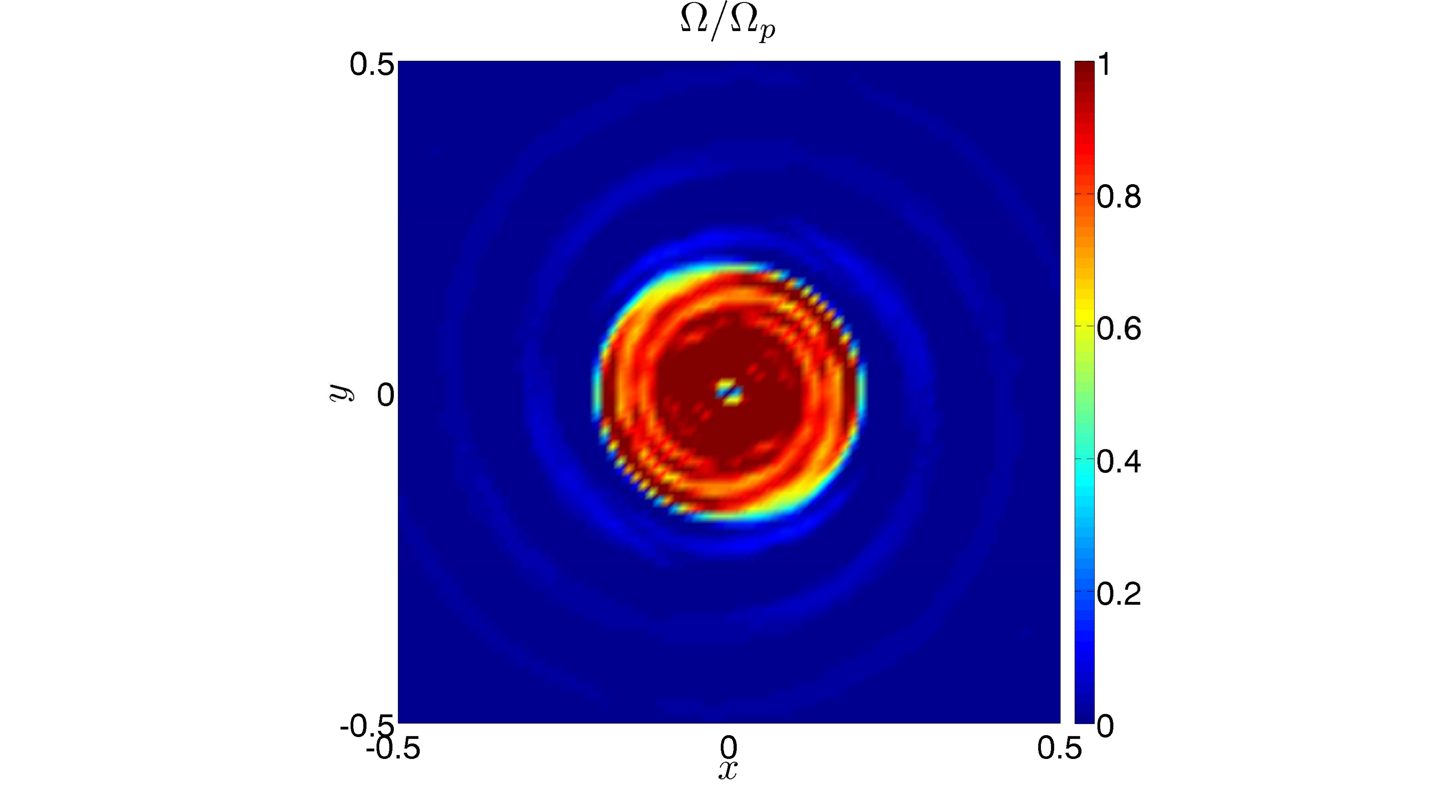} 
    }
    \subfigure{
      \includegraphics[width=0.475\textwidth,clip=true,trim=90mm 1mm 90mm 1mm]{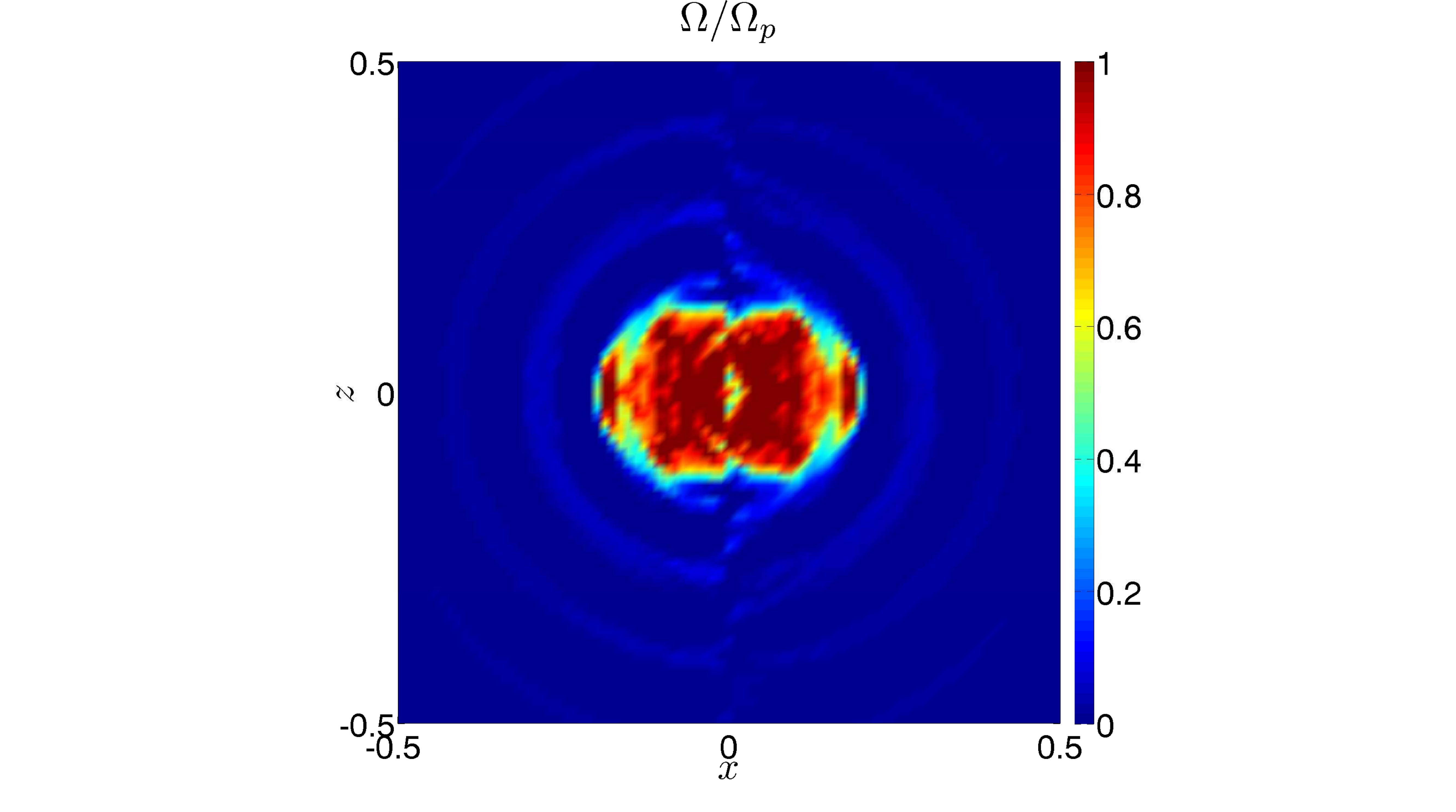} 
    }
  \end{center}
  \caption{2D plot of the angular velocity $u_{\phi}/R$ of the central regions in
    the $xy$ (left) and $xz$ (right) planes, normalised to the angular pattern
    speed (i.e., orbital angular frequency), in a large-amplitude simulation with $\tilde{f}_{r}=1$, at
    $t=450$. Latitudinal differential rotation is produced by
    absorption of ingoing $l=m=2$ waves.
  }
\label{256lam15fr1t45}
\end{figure}

The picture in 3D in the $xy$-plane is very similar to that in 2D, 
as can be seen in Fig.~\ref{256lam15fr1t45xy1} (to compare
with Fig.~6 from BO10). However, one noticeable difference 
is that the primary wave preferentially
transfers its angular momentum at low latitudes, close to the
equatorial plane. This can be seen in Fig.~\ref{256lam15fr1t45}, where
we plot the angular frequency of the flow normalised to $\Omega_{p}$ once a critical layer has
formed, in both the $xy$ and $xz$ planes. This is a consequence of the latitudinal form of
$Y_{2}^{2}$, whose magnitude peaks at $\theta = \pi/2$, as is illustrated in
Fig.~\ref{256lam15fr01t35xz}. The critical layer
absorption is observed to continue as the wave forcing is ongoing, so
this differential rotation is continually reinforced by the absorption
of $l=m=2$ IWs. Since there are wave motions in the region of fluid interior of
the critical layer, parts of these regions spin slightly faster than
$\Omega_{p}$ (this is not seen in Fig.~\ref{256lam15fr1t45} due to the adopted colour
scale). This was also observed in the 2D simulations.

\section{Wave launching region at the top of the radiation zone}
\label{linearisednumerics}

The main motivation for our work is to study $Q^{\prime}$ for
solar-type stars, and in particular to connect this with
the survival of close-in extrasolar planets. 
We have demonstrated that the fate of IGWs approaching the centre of a
solar-type star, outlined in BO10, is unaffected by the extension
to three dimensions. A critical layer is formed by the deposition of
angular momentum through wave breaking once the waves cause
isentropic overturning near the centre. For lower amplitudes, the waves reflect
coherently and approximately perfectly from the centre of the star, and may form global modes. In
that case, the dissipation is only efficient when the system enters a
resonance with a global mode. The corresponding time-averaged
dissipation rate is weak, because the system spends most of its time
out of resonance (T98;
GD98), resulting in negligible tidal evolution of the
planetary companion. Note, however, that this neglects the possibility
that passage through resonance will cause wave breaking, which should be
studied in future work.

If a critical layer forms, wave absorption is efficient,
and global modes (of any frequency very similar to the orbital frequency) 
are prevented from being set up in the RZ. The tidal torque can then
be computed from assuming that the IWs are
entirely absorbed. In this case
a calculation along the lines of GD98 for the ingoing energy flux of
the waves excited at
the top of the RZ is required. This estimates the tidal torque, and thus
the orbital evolution of the planetary companion. In this section, we
perform numerical integrations of the linearised tidal response in an
extensive set of stellar models of solar-type stars with masses in the
range $0.5 \leq m_{\star}/M_{\odot} \leq 1.1$, throughout their main
sequence lifetimes. We aim to determine the tidal torque numerically,
and compare it with a simple model of the launching region at the top
of the RZ, which was derived in GD98 and discussed in BO10.

\subsection{Numerical computation of the linearised tidal response
  throughout the star}

In this section we solve the linearised equations governing the
adiabatic tidal response (Eqs.~\ref{eqmtideeqn} and \ref{dyntideeqn})
throughout the star, computing the excitation of both the equilibrium
and dynamical tides numerically. This allows us to determine the
ingoing energy and angular momentum fluxes in IGWs launched at the top of the RZ, and to
check the validity of approximate semi-analytic formulae for these
quantities, presented in the next section. This is important because the orbital evolution of a
planetary companion is determined by the ingoing angular momentum flux absorbed
at the critical layer. 

We solve the following
coupled ODEs for the radial and horizontal displacements:
\begin{eqnarray}
  \nonumber
\frac{d \xi_{r}}{d r} \hspace{-0.2cm} &=& \hspace{-0.2cm} -\left[\frac{2}{r} +\frac{N^{2}}{g}
  +\frac{d\ln \rho}{dr}\right]\xi_{r} + \left[\frac{l(l+1)}{r} -
  \frac{\omega^{2}r\rho}{\Gamma_{1} p} \right]\xi_{h} \\ && \hspace{3.5cm}
\label{terquem1} + \frac{f r^{2}
  \rho}{\Gamma_{1} p}, \\
  \label{terquem2}
\frac{d \xi_{h}}{d r} \hspace{-0.2cm} &=& \hspace{-0.2cm}\left[\frac{1}{r} - \frac{N^{2}
    }{r\omega^{2}}\frac{d\ln p}{d\ln \rho} \right]\xi_{r} -
\left[\frac{1}{r}-\frac{N^{2}}{g}\right]\xi_{h} - \frac{f N^{2} r}{\omega^{2}g}.
\end{eqnarray}
An outline of the derivation of these equations is presented in T98. 
Note that we are ignoring the self-gravity of the
entire tidal response, which is reasonable because most of the mass of
the star is concentrated near the centre. This assumption is certainly valid for the
dynamical tide, and is approximately valid for the equilibrium tide. 
In these equations, we take the tidal potential in the
frame rotating with $\Omega_{p}$ to be equal to Eq.~\ref{tidpot} with
$\phi=\xi$ and $\omega=0$, so that\footnote{Note that we are defining spherical
harmonics in a standard manner, normalised so that the integral of
$|Y_{l}^{m}|^{2}$ over solid angles is unity, unlike T98.}
\begin{eqnarray}
f = -\sqrt{\frac{6\pi}{5}}\frac{m_{p}}{(m_{\star}+m_{p})}n^{2}.
\end{eqnarray}
This is the amplitude of the largest tide for a
circular orbit. In this frame, the displacement field is separated
into radial and horizontal (non-radial) components
\begin{eqnarray}
\boldsymbol{\xi} =
\xi_{r}Y_{l}^{m}(\theta,\xi)\mathbf{e}_{r}+\xi_{h}r\nabla Y_{l}^{m}(\theta,\xi).
\end{eqnarray}
The tidal response is further decomposed into an equilibrium and a dynamical
tide, as defined in the introduction.

We impose a free upper boundary, i.e., take the Lagrangian pressure
perturbation $\Delta p = 0$ at $r=R_{\star}$, so that the Eulerian pressure perturbation
$\delta p = - \xi_{r} \frac{dp}{dr}$.
Since the relation 
\begin{eqnarray}
\label{nonradialeom}
\xi_{h} = \frac{1}{\omega^{2}r}\left(\frac{\delta p}{\rho} + fr^{2} \right),
\end{eqnarray}
follows from the non-radial equation of motion, this relates our variables at the
surface of the star. We take an IW BC at the inner boundary
at $r \approx 0.02 R_{\star}$, where we match $\xi_{r}$ and $\xi_{h}$ onto the analytic
solution for an IW derived in \S \ref{lineartheory3D},
\begin{eqnarray}
  \xi_{r}(r) &=&  A_{\xi} r^{-\frac{3}{2}}(J_{\frac{5}{2}}(k r)+ i Y_{\frac{5}{2}}(k r)), \\
 \nonumber \xi_{h}(r) &=& \frac{A_{\xi}}{6}r^{-\frac{3}{2}}\left[3(J_{\frac{5}{2}}(k r)+ i Y_{\frac{5}{2}}(k r))
    \right. \\ &&\left. \hspace{2cm} -kr(J_{\frac{7}{2}}(k r)+ i Y_{\frac{7}{2}}(k r))\right],
\end{eqnarray}
where $k = \frac{m C}{\omega}$. We take $A_{\xi}$ to be a free
parameter, so that these equations relate the ratio of $\xi_{r}$ and $\xi_{h}$. This solution is quite accurate when
$r/R_{\star} \lesssim 5\%$ since $\boldsymbol{\xi}^{e}$ is negligible
in this region. This BC is meant to represent
the IW absorption at a critical layer.

We solve Eqs.~\ref{terquem1}--\ref{terquem2} using data interpolated
from a stellar model at points required by a 4th/5th order adaptive
step Runge-Kutta integrator, using a cubic spline interpolation. In
particular, the coefficients of $\xi_{r}$, $\xi_{h}$, in
Eqs.~\ref{terquem1}--\ref{terquem2} are
singular at the origin, so we first multiply these quantities by $r$ \textit{before} interpolating
their values to the locations required by the ODE integrator, using
the stellar model parameters. We then divide by $r$, after the
interpolation. This is done to get the correct behaviour for small $r$.

Our method of solution is a shooting
method to an intermediate fitting point (\citealt{Press1992}), which we take to be the CZ/RZ
interface, where we enforce continuity of the solution.
The freely specifiable initial conditions for each ODE integration are chosen to
be $\xi_{r}$ at the surface, and $A_{\xi}$ at the inner boundary. We use
$\xi_{r}=\xi_{r}^{e}$ as our starting ``freely specifiable'' estimate
at the surface, which is an accurate approximation
since $|\xi_{r}^{d}| \ll |\xi_{r}^{e}|$ at $r=R_{\star}$.

The Eulerian pressure perturbation for the dynamical tide is
\begin{eqnarray}
\delta p^{d} = \rho
\omega^{2} r \left(\xi_{h}^{e} + \xi_{h}^{d}\right),
\end{eqnarray}
from the horizontal component of Eq.~\ref{dyntideeqn}. The radial energy flux at each radius is
\begin{eqnarray}
F = \frac{\omega r^{2}}{2} \mathrm{Im}\left[(\delta p^{d})^{*}
  \xi^{d}_{r}\right],
\end{eqnarray}
which follows from manipulating Eqs.~\ref{eqmtideeqn} and
\ref{dyntideeqn} to derive an
energy equation.

From Eq.~\ref{dyntideeqn}, we derive the relation
\begin{eqnarray}
\nabla \cdot \mathrm{Im}\left\{ \delta p^{d} (\boldsymbol{\xi}^{d})^{*}
\right\} = \rho
\omega^{2}\mathrm{Im}\left\{(\boldsymbol{\xi}^{d})^{*}\cdot \boldsymbol{\xi}^{e}  \right\},
\end{eqnarray}
which can be averaged over solid angle to give
\begin{eqnarray}
\nonumber
&& \mathrm{Im}\left\{\partial_{r}\left(r^{2} \delta p^{d} (\xi^{d}_{r})^{*} \right)\right\} = \rho r^{2} \omega^{2}
\mathrm{Im}\left\{(\xi^{d}_{r})^{*}\xi^{e}_{r} \right. \\ &&
\hspace{4.2cm} \left. + l(l+1)(\xi^{d}_{h})^{*}\xi^{e}_{h}  \right\}.
\end{eqnarray}
This is telling us that the dynamical tide is
forced by the equilibrium tide. As part of the validation of our numerical code, we have confirmed
that this is accurately satisfied from the numerical solutions
computed with an ingoing BC. This should adequately convince ourselves that the
code is able to accurately compute the energy flux in the dynamical tide.

\subsection{Results}

\begin{figure}
  \begin{center}
    \subfigure{
      \includegraphics[width=0.48\textwidth]{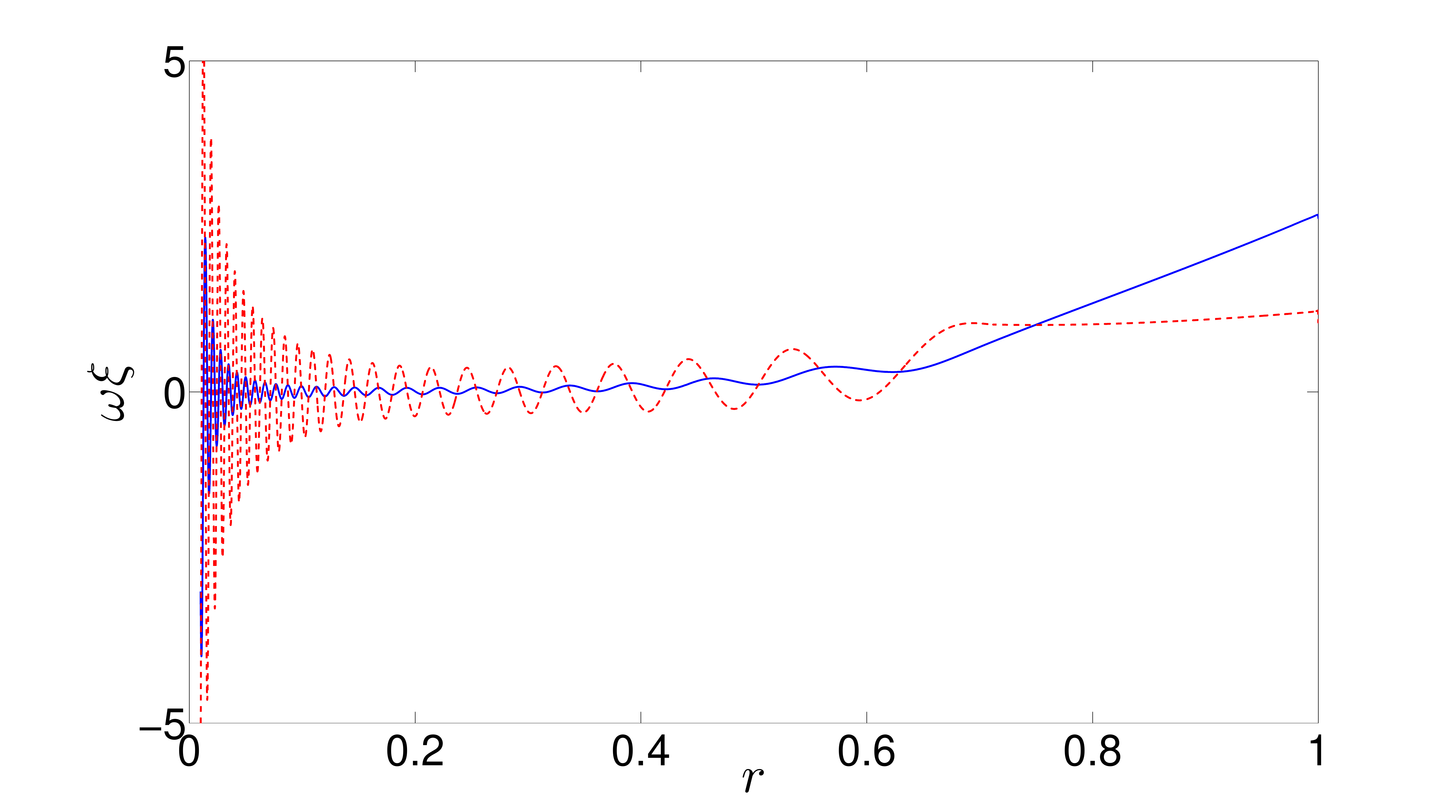} }
  \end{center}
  \caption{Typical values of the real parts of the radial ($\omega \xi_{r}$, in blue solid
     lines) and horizontal ($\omega \xi_{h}$, in red dashed lines)
     velocity components, in $\mathrm{m}\,\mathrm{s}^{-1}$. We use Model S of the
     current Sun, and consider the tidal perturber to be a $P=1$ d,
     $m_{p} = 1 M_{J}$, planet orbiting the current Sun.}
  \label{P1Mp1TerquemIngoingBC}
\end{figure}

As an illustration, we present the results of our integrations for a
fiducial case with a $P=1$ d, $m_{p}=1 M_{J}$ planet orbiting the
current Sun (for which we use Model S, described in \citealt{CD1996}) in
Fig.~\ref{P1Mp1TerquemIngoingBC}. We plot the real parts of $\omega
\xi_{r}$ (solid blue) and $\omega \xi_{h}$ (dashed red) throughout the
star, which represent typical values of the radial and horizontal
velocity components. The radial wavelength of the waves decreases as the waves
propagate deeper into the RZ, where $N^{2}$ increases. The large increase in the velocity
amplitude near the centre is evident, as predicted from the linear
solutions in \S~\ref{lineartheory3D}. This can be compared with a
similar calculation in T98, displayed in their Fig.~1, which our code
correctly reproduces for the given planetary orbital period when a regularity
condition in applied at the centre. The only difference between our
calculations and theirs is that we use an IW BC, whereas they allow
the waves to perfectly reflect from the centre.

\begin{figure}
  \begin{center}
    \subfigure{
      \includegraphics[width=0.48\textwidth]{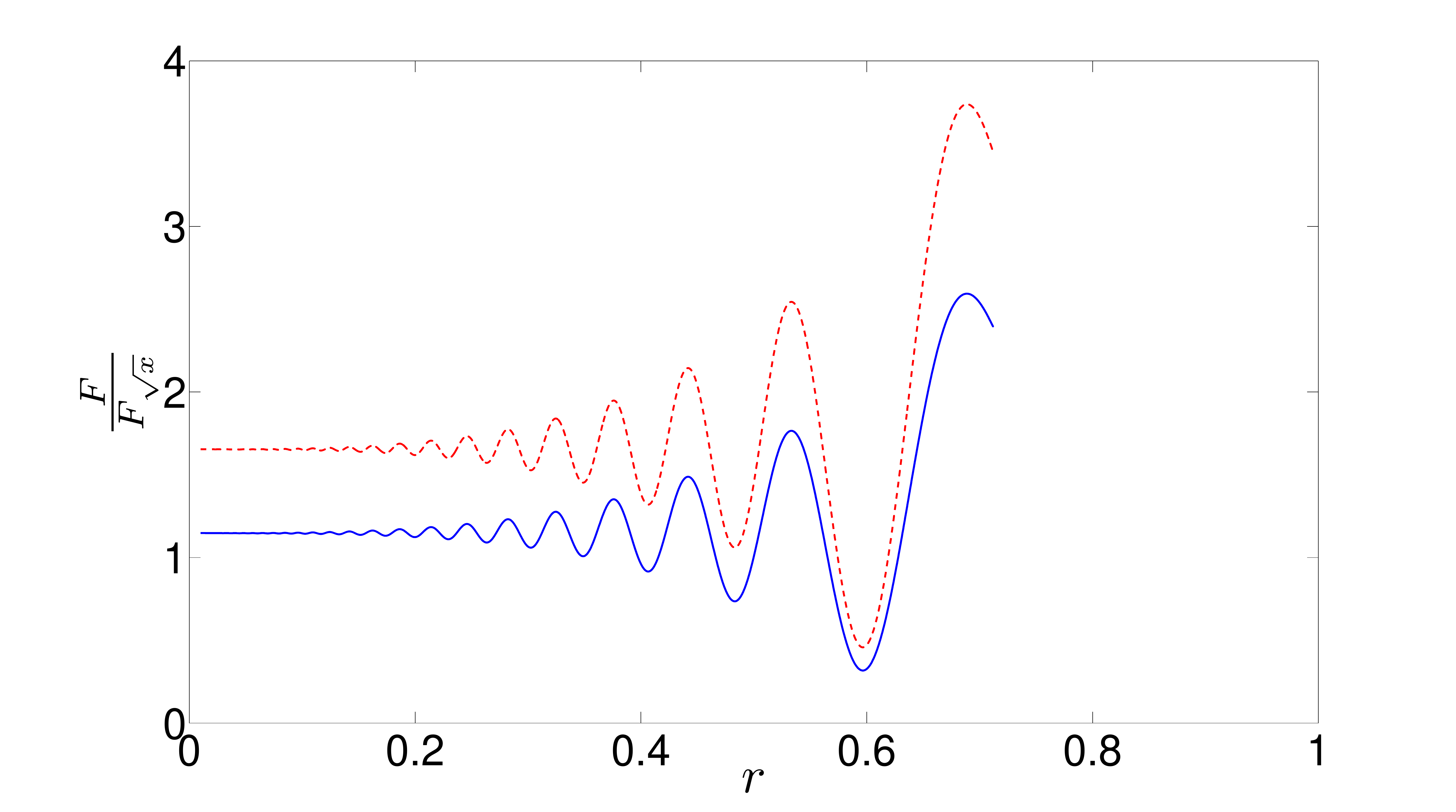} }
   \end{center}
  \caption{Numerically computed $F$, normalised to the
    prediction from the analytic theory in Eq.~\ref{besselenergyflux}
    (solid blue line) and GD98 Eq.~13 (red dashed
    line). Eq.~\ref{besselenergyflux} overestimates $F$ by less
    than $10\%$, and is thus a good approximation of the ingoing
    energy flux.}
  \label{FnumP1Mp1TerquemIngoingBC}
\end{figure}

In Fig.~\ref{FnumP1Mp1TerquemIngoingBC} we plot the ingoing energy flux,
normalised to both the semi-analytic prediction of GD98 (red dashed
lines), and a revised expression derived in Appendix
\ref{energyfluxcalculation} (blue solid lines), which will be discussed in the next
section. This illustrates that the ingoing energy flux
oscillates about its final asymptotic value, which it eventually approaches deep
in the RZ.

\subsection{Semi-analytical calculation of the ingoing energy flux}

In this section we compare the numerically computed $F$ with the
semi-analytic estimate of GD98, and present a slight
refinement which is found to be appropriate for solar-type stars.

 \begin{figure}
   \begin{center}
     \subfigure{
       \includegraphics[width=0.48\textwidth]{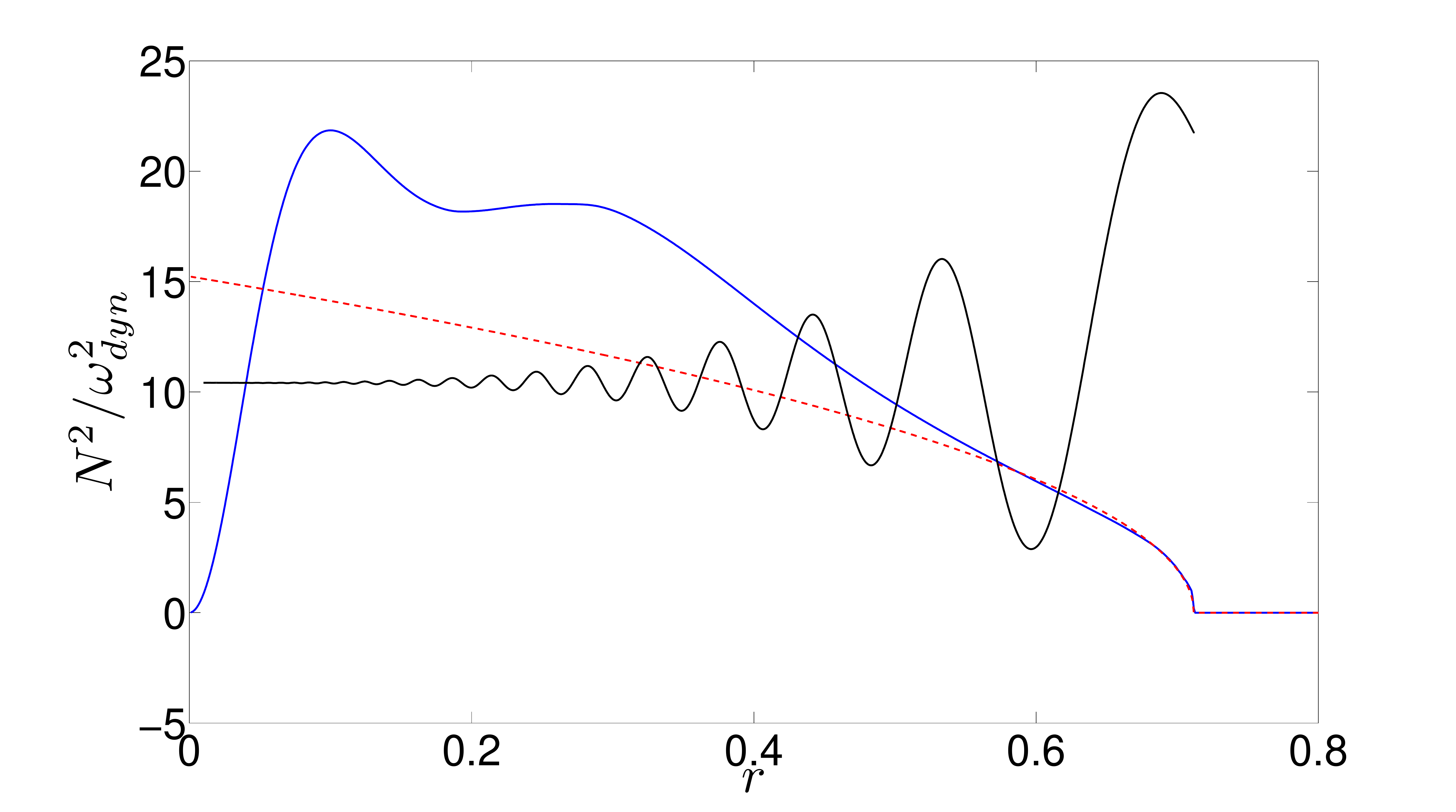} }
   \end{center}
   \caption{$N^{2}$ normalised to $\omega^{2}_{dyn}$ (solid
     blue line), together
     with our local approximation $N^{2} \propto
    (r_{b}-r)^{1/2}$ (red dashed line). This approximation is 
     reasonable over a region of size $\sim 0.15R_{\odot}$. 
     We also plot an arbitrarily scaled profile of $F$, 
     to compare its radial profile with the profile
     of $N^{2}$, for $P = 1$ d.}
   \label{N2vsrSqrtxOverlay}
 \end{figure}

 In the launching region at the top of the RZ, there is a location at
 which $N^{2} = 0$ at $r=r_{b}\approx 0.71 R_{\odot}$, near to which it follows from the
 dispersion relation (Eq.~4 in BO10) that the radial wavenumber of
 gravity waves vanishes. Near this turning point we can approximate
 the solution in this region by assuming a functional form for
 $N^{2}$. GD98 take $N^{2} \propto r_{b}-r \equiv x$, in which case the problem
 in the launching region reduces to the solution of Airy's
 differential equation for $\xi_{r}^{d}$. In this case GD98 derive
 their Eq.~13 for the resulting ingoing energy flux (Eq.~92 in BO10,
 which we denote herein as $F_{x}$).
 This should match the asymptotic numerical computation of
 $F$. We find that a slightly better
 approximation is to take $N^{2} \propto \sqrt{x}$ over the
 launching region\footnote{However, no simple physical arguments for
   such a profile have been found, which is perhaps to be expected
   since stellar models contain complicated combinations of physics.}, since this is valid over $\sim 0.15R_{\odot}$ from
 the interface, as we illustrate in Fig.~\ref{N2vsrSqrtxOverlay}. The
 slope of the curve $\sqrt{x}$ can be obtained through fitting
 to the profile of $N^{2}$ in the stellar model. This allows a
 slightly more
 accurate calculation of $F$ based on the stellar model than is
 obtained through direct application of $F_{x}$ (where the gradient
 $d N^{2}/ d x$ is not uniquely defined). Nevertheless, the
 differences are only a factor of two at most. In Appendix
 \ref{energyfluxcalculation} 
 we calculate the solution in the launching region and the resulting $F$, for
 any power law profile of $N^{2}$ with a positive exponent, using the
 framework of OL04. Using our approximation, the radial extent of the launching
 region $L_{launch} \sim 0.03 R_{\odot}$. 

 For the case $N^{2} \propto \sqrt{x}$, we obtain, for a non-rotating background, 
\begin{eqnarray}
  \label{besselenergyflux}
\nonumber
F \equiv F_{\sqrt{x}} &=& \frac{\left(\frac{2}{5}\right)^{\frac{1}{5}}\pi}{2
  \left[\Gamma\left(\frac{3}{5}\right)\right]^{2}}
\left[l(l+1)\right]^{-\frac{7}{5}}\omega^{\frac{19}{5}} \\ &&
\hspace{1.4cm} \times \left[\rho_{b}
  r_{b}^{5}  \left|\sqrt{r_{b}}\frac{d N^{2}}{d \sqrt{x}}\right|^{-\frac{2}{5}}
  \left|\frac{\partial \xi_{r}}{\partial x}\right|^{2}\right].
\end{eqnarray}
We express $\partial_{x} \xi_{r} =
\sigma_{c} \Psi/\omega_{dyn}^{2}$, and compute $\sigma_{c}$ by solving
GD98 Eq.~3 in the CZ, using a linear shooting method, with the boundary conditions that
$\xi_{r}^{d}(r_{b}) = \xi_{r}^{d}(R_{\odot})= 0$.
This matches the numerically computed asymptotic value of $F$ for the
current Sun quite well, to within $10\%$ (see
Fig.~\ref{FnumP1Mp1TerquemIngoingBC}). The remaining discrepancy is
due to the slight variation in 
background density, and the magnitude of the equilibrium tide, over the launching region. We note
that the main change in the energy flux from modifying the profile of
$N^{2}$ in the transition region is to change its frequency
dependence.

As can be seen from Fig.~\ref{FnumP1Mp1TerquemIngoingBC}, the
 numerically computed asymptotic value of $F$ differs from $F_{x}$ by a factor
 $\lesssim 2$, even in the case of short-period forcing. We would
 expect $F_{x}$ to be correct to within a factor $\sim 2$ even
 if the slope varies by an order of magnitude, since it is only raised to
 the power $-\frac{1}{3}$. However, our slight refinement improves this estimate by $\sim 50\%$ for $P =
 1$ d.

We have performed numerical integrations for several different forcing
frequencies (planetary orbital periods), for which we find $F \propto
P^{-8.06}$ for fixed stellar and planetary properties (other than the
orbital period). If we take into account the fact that
$\partial_{x}\xi_{r}^{d} \sim
\Psi \sim P^{-2}$, and then consider a fixed tidal potential, we find that $F \propto \omega^{4.06}$, which
is slightly different from the power law dependence in
Eq.~\ref{besselenergyflux}. The discrepancy most likely results from
the variation in the background density, and the magnitude of the
equilibrium tide (which forces the dynamical tide), within this
region.

For $P \lesssim 2 \; \mathrm{d}$, which is the
relevant regime for which this process is potentially important for
the survival of close-in planets (BO10), there is a few per-cent variation in
background parameters over a lengthscale $L_{launch}$. This means that our calculation of
$F_{\sqrt{x}}$ differs from the numerically computed value of $F$, by an amount that
increases as $P$ is made smaller to a maximum of $20\%$ when $P=0.5$ d.
Nevertheless, this discrepancy is small, and our semi-analytic
estimate $F_{\sqrt{x}}$ is a good approximation to $F$ for all cases
that we have modelled. This is used to provide an estimate of
$Q^{\prime}$, and thus the tidal torque, in \S~\ref{Qandexplanation}.

\subsection{Variation between different stellar models}
\label{VariationBetweenStellarModels}

Eq.~\ref{besselenergyflux}, matches the numerically computed value to within a few per-cent 
for a variety of solar-type stars. This is because it generally arises
that $N^{2}\propto x^{\alpha}$, with $\alpha \sim 0.5$, when the launching region
is a few percent of the stellar radius. This occurs for the waves excited by
planets in short-period orbits. We have confirmed this by computing
$F$ in a number of stellar models with masses in the range $0.5 \leq
m_{\star}/M_{\odot} \leq 1.1$, and ages that represent the range of
main-sequence ages expected for these stars. 
These were computed using ASTEC (\citealt{JCD2008}). 

The main uncertainty in these models is the profile of $N^{2}$ within
the launching region, especially since the slope within $0.02 R_{\odot}$ of the interface is not well constrained by
theory or observations. Helioseismic observations are not yet able
to constrain the stratification within this
region, owing to the lack of observed g-modes
(\citealt{EllisGough1984}; \citealt{QuestSolargmodes2009}). In addition, the relevant
physics included in the stellar models is also uncertain, particularly
as a result of changes to the compositional gradient from convective overshoot. The
inclusion of helium settling tends to make the interface
profile sharper, and the inclusion of turbulent diffusion, that results from
convective overshoot, and often parameterised using a simplified 1D model of
this process, tends to smooth out the profile near the
interface (J\o rgen Christensen-Dalsgaard, private communication). However, these changes are
small and occur only within a region smaller than the launching region for $P\lesssim 3$ d. This means that
uncertainties in the observations, and the physics, at the interface are
unlikely to significantly change $F$ for $P\lesssim 3$ d, which are
the planets whose survival could be threatened by our mechanism (as we
discuss below). In addition, $\left|\frac{dN^{2}}{d
  \sqrt{x}}\right|$ is only raised to the -2/5 power in our model. 

\begin{figure}
   \begin{center}
     \subfigure{
       \includegraphics[width=0.47\textwidth]{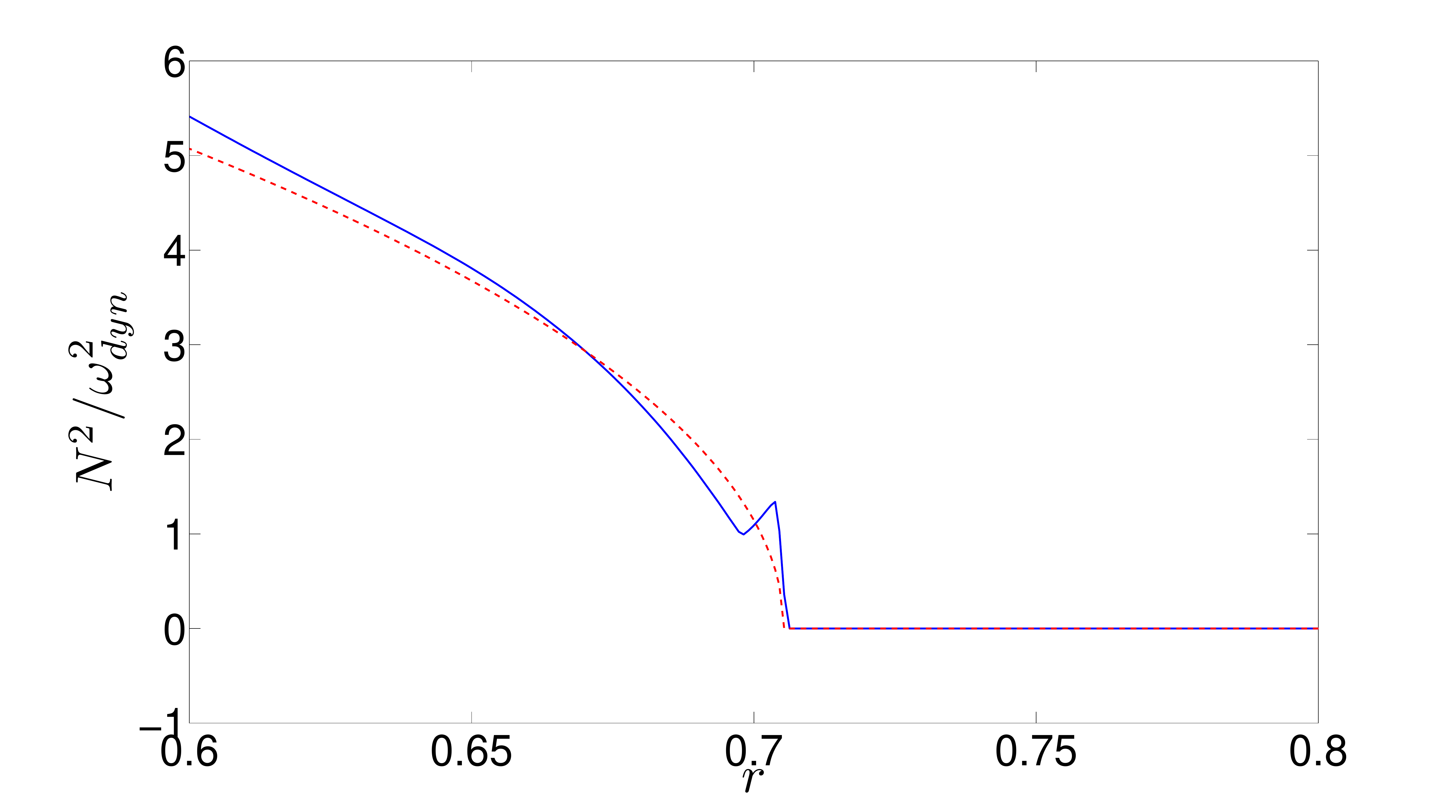} }
      \subfigure{
        \includegraphics[width=0.47\textwidth]{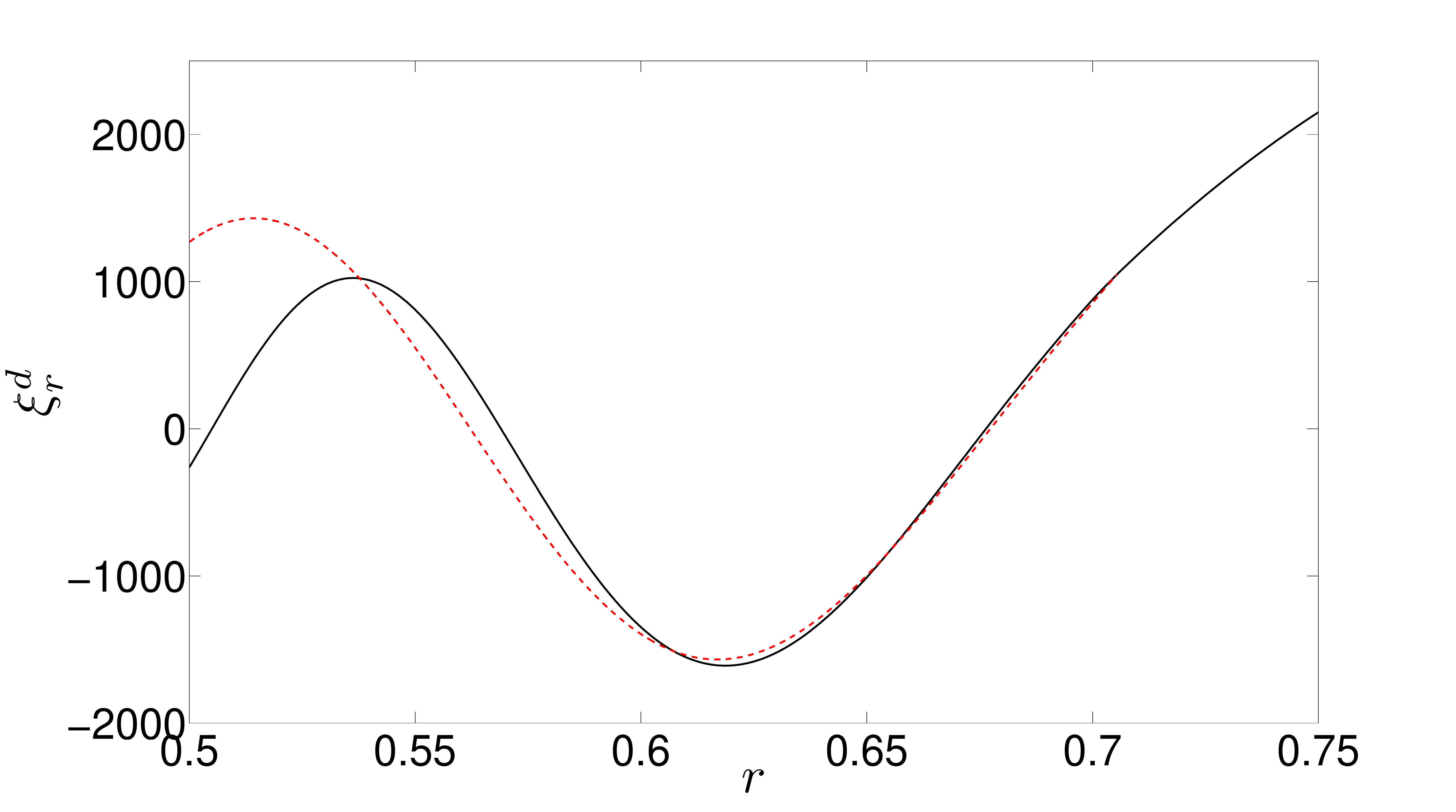} }
   \end{center}
   \caption{In the top panel we plot $N^{2}$ in a $1.0M_{\odot}$ star
     at $t=6.95$ Gyr, with our fitted solutions in the bottom
     panel. This model has a ``bump'' in the $N^{2}$ profile, but this
     occurs over a region smaller than $L_{launch}$, and so does not
     reduce the accuracy of our analytic model
     (from Eq.~\ref{besseltransitionsolution}) in the launching
     region (or the corresponding energy flux), shown in the bottom panel.
   }
   \label{DifferentStellarModels}
 \end{figure}

For Model S, the compositional gradient
($\nabla_{\mu}$) is unimportant in the launching region, and $N^{2}$ primarily results from temperature
gradients. As the star evolves, the settling of elements heavier than $H$,
produces a compositional gradient at the top of the RZ, which can
produce ``bumps'' in the profile of $N^{2}$. In Fig.~\ref{DifferentStellarModels} we plot an example of the profile
of $N^{2}$ for a $1 M_{\odot}$ model (with $Z=0.02$) at $t=6.95$ Gyr,
together with the best fit solution in the launching region. For $P\lesssim 3$ d, $L_{launch}$ is
generally much larger than these ``bumps'', so the wave launching process does
not notice such departures from a smooth stratification profile, and
Eq.~\ref{besselenergyflux} remains a good approximation for the energy
flux. However, if the frequency is sufficiently low ($P\gtrsim 6$ d), the radial extent
of the launching region can become comparable with the size of these
``bumps'', and the numerical solution can depart appreciably
from our analytical model. Planets in such orbits are very unlikely to be
affected by stellar tides at such orbital distances, so we do not
consider such effects worthy of further consideration. To summarise,
we have confirmed that Eq.~\ref{besselenergyflux} is a good
approximation for the energy flux for planets on orbits of a few
days throughout the range of solar-type stars in our study.

\subsection{Are 1D linear hydrodynamic calculations reasonable?}

The calculations done in \S~\ref{linearisednumerics} are performed under the
assumption of linearity. We have demonstrated in \S~\ref{3dresults}
and in BO10, that this is not a valid
assumption near the centre of a star. A simple
estimate of the nonlinearity in IGWs in the launching region, for want of a better
measure, compares the radial
displacement to the radial wavelength, for which $\xi_{r}^{d}/\lambda_{r} \sim 10^{-4}$,
for a hot Jupiter orbiting the Sun on a one-day orbit (this estimate can be obtained
from Fig.~\ref{P1Mp1TerquemIngoingBC}). This value increases
linearly with the mass of the planet, but we are still in the linear regime
even if we have a close-binary perturber, which indicates that
linearity is likely to be a good approximation in the launching region
(and throughout the RZ, except for the central regions).

However, it remains to be seen whether $F$ for the 1D calculations is the same
as that in 2D or 3D numerical simulations of realistic tidal forcing
in a model including both a CZ and a RZ. Such
simulations as \cite{Rogers2006a} could be performed of the whole star subject
to tidal forcing. The turbulent convection in these simulations will
produce a spectrum of waves at the top of the CZ, in
addition to those excited by tidal forcing. It may be that the
interaction of these waves reduces the ingoing energy
flux. Alternatively, the profile of $N^{2}$ in the transition region
could be modified by realistic modelling of convective overshoot, altering the strength of
the stratification within a few percent of a pressure scale height from
the interface. This would affect $F$ if the overshoot region is
comparable with the size of the launching region, though probably not
significantly, as we discussed in the previous section.

A toroidal magnetic field in the launching region could also affect the amplitudes of
these waves, and the value of $F$. \cite{Rogers2010} find
that when the IGW frequency is approximately equal to the Alfv\'{e}n
frequency ($\omega_{A}$), strong wave reflection occurs. This behaviour
follows from the dispersion relation for IGWs in the presence of a
magnetic field (e.g.~\citealt{Kumar1999}), and could have important
consequences for $F$, if the magnetic field is sufficiently
strong. However, for $\omega \approx \omega_{A}$ in the launching region, we require
$B_{\phi} = 2\pi\sqrt{\mu_{0} \rho_{b}}r_{b}/P \gtrsim 6 \;
\mathrm{MG}$, when $P=1$ d, which is close to being ruled out on empirical grounds, from
measurements of the solar oblateness (\citealt{Friedland2004}).
If $\omega_{A} < \omega$, little attenuation of wave energy over the
non-magnetic case is found by \cite{Rogers2010}, so this seems unlikely
to affect $F$ in our case.

In the innermost wavelength, a strong magnetic field would be able to
reflect IGWs before they reach the centre, if $\omega \approx
\omega_{A}$ in this region. This requires a toroidal field of strength
$B_{\phi}  \gtrsim 6 \; \mathrm{MG}$, or a poloidal (radial) field of strength
$B_{r} \gtrsim 28 \;\mathrm{MG}$. Since it is unlikely that such
fields could exist in the RZ \citep{Friedland2004}, 
a magnetic field will probably not affect the reflection of IGWs
(excited by short-period planets) in
the RZ. It is therefore appropriate to ask whether the waves will break on
reaching the centre.

\section{Tidal $Q^{\prime}$ and the breaking criterion}
\label{Qandexplanation}

The modified tidal quality factor of the star that results from critical layer
absorption can be obtained from Eq.~\ref{besselenergyflux} (a slight
modification from BO10, resulting from the adopted profile of $N^{2}$
in the launching region),
evaluating to
\begin{eqnarray}
\label{Qfactorcalc}
  Q^{\prime} \approx 9 \times 10^{4} \left[\frac{P}{1 \;\mathrm{day}}\right]^{\frac{14}{5}},
\end{eqnarray}
with the exact value depending on the stellar model adopted.
However, this value is found to vary by only a
factor of 5 throughout the range of main-sequence stars in our study,
for a given orbit. As the star evolves on the main-sequence, the position of the
interface $r_{b}$ moves inwards towards higher density material, which
slightly increases $F$. This means that $Q^{\prime}$ tends to
decrease with main-sequence age, though only by up to a factor of 5.
The dissipation that results from wave absorption at a critical layer
thus becomes more effective as the star evolves.

The waves break if
\begin{eqnarray}
\label{breakingcriterion}
\beta\left(\frac{C}{C_{\odot}}\right)^{\frac{5}{2}}\left(\frac{m_{p}}{M_{J}}\right)\left(\frac{M_{\odot}}{m_{\star}}\right)\left(\frac{P}{1 
\;\mathrm{day}}\right)^{\frac{1}{10}} \gtrsim 3.6,
\end{eqnarray}
where $1/2 \leq \beta \leq 2$ is a coefficient\footnote{Which was written as
$(\tilde{\mathcal{G}}/\tilde{\mathcal{G}}_{\odot})^{1/2}$ in BO10
Eq.~103; see that paper for further details.} which is unity for the
current Sun but varies with the stellar model, and $C_{\odot} = 8\times 10^{-11}
\mathrm{m}^{-1}\mathrm{s}^{-1}$. This expression is valid if we are
sufficiently far from resonance with a global mode. Note, however,
that if we are close to a resonance, the central amplitude may be
much larger than these estimates would predict, which would make breaking more
likely. Note also that this criterion becomes much easier to satisfy
as the star evolves, since $C$ increases with main-sequence age (see
Fig.~1 in BO10). If a planet exceeds this criterion, then its orbital
migration is rapid ($\sim \mathrm{Myr}$; see Eq.~105 in BO10) once
critical layer absorption occurs, and would therefore threaten the
survival of any short-period planet that causes wave breaking with
$P\lesssim 3$ d. Very
massive perturbers, with larger orbital moments of inertia than the
spin moment of inertia of the RZ of the star, may be able to
synchonise the spin of the RZ with the orbit, and prevent further
inward migration. This may be an explanation for the synchronisation
of close-binary stars (e.g.~\citealt{Mazeh2008}).

\subsection{The survival of short-period planets}

Our most important result is that this mechanism can potentially
explain the survival of all short-period extrasolar planets around
solar-type stars\footnote{Contained in the catalogue at
  http://www.exoplanet.hanno-rein.de/ or
  http://exoplanet.eu/catalog.php} with masses in the range $0.5
\leq m_{\star}/M_{\odot} \leq 1.1$. From using the closest fit stellar
model to each of these stars, we find that no planet clearly satisfies 
Eq.~\ref{breakingcriterion} with a period $P \lesssim
3$ d (planets much further out may satisfy the criterion, though this tidal
effect is then unimportant). All of these planets
are insufficently massive (or orbit sufficiently young stars, with low
values of $C/C_{\odot}$) that
they are unlikely to cause wave breaking at the centre of their
hosts. The dominant mechanism of tidal dissipation in these stars is
therefore likely to be damping of the equilibrium tide by turbulent
convection. This is likely to be relatively inefficient for these periods, since the
turbulent viscosity must be reduced when the orbital period is shorter
than the convective timescale (see e.g.~\citealt{Zahn1966};
\citealt{Goldreich1977}; \citealt{GoodmanOh1997};
\citealt{Penev2007}). This could be an important explanation for the
survival of these planets. 

This mechanism can also explain why the most massive short-period
planets (which still have smaller or comparable moments of inertia in the orbit
as the RZ of the star), such as WASP-18 b \citep{Hellier2009}, WASP-14
b \citep{Joshi2009} or
CoRoT-14 b, are exclusively found around F-stars, which have
convective cores, and in which critical layer formation induced by
wave breaking at the centre is unable to
operate. In addition, these stars have been found to have weaker tidal dissipation
in their outer CZs than G-stars \citep{Barker2009}. Note, however, that planets with much larger masses may have sufficient orbital
angular momentum to be able to synchronise their stars and reach an
approximate tidal equilibrium state, neglecting stellar magnetic
braking, which would prevent orbital decay (e.g.~\citealt{Levrard2009}), even for planets in orbit around a
solar-type star. Consequently, we make the prediction that fewer massive planets in the range $3 \lesssim
M_{J} \lesssim 20$, in orbits with $P \lesssim 2-3$ d, that satisfy
Eq.~\ref{breakingcriterion}, will be found around solar-type stars.

\section{Conclusions}

In this paper, we have presented the results of a set of 3D
simulations of IGWs approaching the centre of a solar-type star. The
aim of this work is to determine the importance of a nonlinear
mechanism of tidal dissipation in solar-type stars, first proposed by
GD98, continuing the investigation of BO10. We have confirmed that the main
results of BO10 are unaffected by the extension to 3D by performing
numerical hydrodynamical simulations, using a Boussinesq-type model
solved with a pseudospectral method.

We first derived a linear wave
solution in 3D, and found that nonlinearities do not vanish for this wave, unlike
the 2D solution, which is exact. Nevertheless, these waves are
found to reflect approximately perfectly for moderate-amplitudes, a
result which we have qualitatively confirmed in numerical
simulations of moderate-amplitude tidal forcing.

The general picture for high-amplitude forcing is that IGWs break within the innermost
wavelengths of a star, if they reach the centre with sufficient
amplitude to overturn the stratification. If this occurs, they form a
critical layer, which we have confirmed from the simulations
efficiently absorbs ingoing wave angular momentum. This
results in the star being spun up to the orbital angular frequency of the planet, from the
inside out. This could be very important to the survival of massive planets in short-period
orbits around solar-type stars.

One noticeable difference in 3D is that the absorption of $l=m=2$
ingoing waves results in the formation of latitudinal differential rotation. This is
perpetually reinforced by critical layer absorption. Instabilities may
act on this rotation profile, which could homogenise the horizontal
angular momentum distribution. These include shear instabilities,
which can be linear \citep{Watson1981} or nonlinear instabilities,
that set in at a critical Reynolds number \citep{RichardZahn1999}.
These have growth times comparable to the tidal period, and could
transfer angular momentum latitudinally. There are also doubly
diffusive instabilities (\citealt{GSF1967}; \citealt{Knobloch1982}),
or magnetic instabilities, such as the magnetorotational instability
\citep{Balbus1994}. However, these mechanisms are unlikely to be able to prevent
the critical layer absorption, and thus prevent the tidal engulfment
of a short-period planet.

In these simulations we artificially forced waves at the top of the
computational domain. In reality, the waves are launched
at the top of the RZ. In this paper we also solved the linearised
equations for the adiabatic tidal response, in order to study the
influence of the launching region in more detail. We provide an
expression for the energy flux Eq.~\ref{besselenergyflux}, which has
been checked to agree very well with the numerically computed energy
flux for all orbits for which this effect is potentially important, 
and in a wide range of solar-type stellar models. We then presented the
breaking criterion that must be satisfied for the waves to break,
slightly refining BO10.

Finally, in \S~\ref{Qandexplanation} we presented the tidal
$Q^{\prime}$ that results from this process, together with a breaking
criterion that determines when these waves break. This allowed us to outline a possible
explanation for the survival of all short-period extrasolar planets,
that is not in conflict with current observations, and makes
predictions which will be tested by ongoing observational studies such
as WASP and Kepler.

As in BO10, our simulations do not show any instabilities that act on the waves when
they have insufficient amplitude to overturn the stratification. This
result suggests that to obtain any efficient tidal dissipation, the
waves must have sufficient amplitude to satisfy
Eq.~\ref{breakingcriterion}, so that they overturn the
stratification. However, it may be that weaker parametric
instabilities operate for waves with lower amplitudes (suggested
by GD98). A detailed stability analysis of the 2D exact wave solution
written down in BO10 is ongoing, and should shed light on this matter.

In our explanation in the previous section, we ignored the potential
effects of the evolution of either the stellar eigenfrequencies or the tidal
frequency, that could result in a passage through a resonance with a global
mode of oscillation. If this occurs, and simple estimates suggest that
this is very likely at least once in the lifetime of a system, then
the waves may break, even if the amplitude of the tide is small before
the passage through resonance. Future work is required to study this problem in
detail.

The simulations performed in this paper used a Boussinesq-type model,
which is only valid in the central $\lesssim 5 \%$ by radius of a
solar-type star. For future work, it is possible to perform simulations using an anelastic code, which could
take into account the radial variation in stellar
properties throughout the whole RZ (e.g.~\citealt{Rogers2005}). We
have also neglected the rotation (or possible differential rotation)
of the RZ, which is reasonable since we are considering the waves
excited by short-period planets in slowly rotating stars. However, the
inclusion of rotation (and differential rotation) would certainly be
interesting to study for more rapidly rotating stars. One consequence
could be that the latitudinal differential rotation observed
in the simulations in 3D may be enhanced by rotation, because
inertia-gravity waves can be confined in the equatorial regions.

In addition, we have so far considered a planet on a circular,
coplanar orbit. Since many planets have been observed with eccentric
or inclined orbits, this makes a study of the circularisation and
potential spin-orbit alignment through wave breaking and the resulting
critical layer absorption, seem an attractive avenue of future research.

\section*{Acknowledgments}
I would like to thank STFC for a research
studentship, Gordon Ogilvie for his insight and for many useful
discussions, J\o rgen Christensen-Dalsgaard for providing a set of
stellar models, and finally the referee for several helpful suggestions which have
improved the readability of the manuscript.

\appendix

\section[]{IW/OW decomposition}
\label{refcoeff}

In \S~\ref{lineartheory3D} we derived an analytic solution for the
waves near the centre. We can use this to deconstruct the numerical solution
into a single IW and OW, as in BO10. Doing this enables us to
quantify the amount of angular momentum absorbed as these waves approach and/or
reflect from the centre. However, the reduction in
resolution per dimension required for the transition to three
dimensions means that an alternative method is preferable, which
includes the information at several points, to reduce numerical
error. The approach we use is now described.

At every point with spatial coordinates $(x,y,z)$,
we have four pieces of information, namely, $u_{r}$, $u_{\theta}$,
$u_{\phi}$ and $b$. Thus, it is possible to compute the (complex) IW
and OW amplitudes $A_{in}$ and $A_{out}$ for the $l=m=2$
wave at a single point. This is, however, computationally expensive, since the routines for computing Bessel
functions are relatively slow (typically several hundred times slower than a square
root). In addition, this method would be subject to potentially significant
round-off errors.

We fit the simulation output to a linear model,
corresponding to a an IW and an OW. This problem is an overdetermined
system of linear equations in terms
of the unknown wave amplitudes, which we write as
\begin{eqnarray}
y_{i} = \sum_{i=1}^{4M} A_{ij} x_{i},
\end{eqnarray}
where $\mathbf{y}=(u_{r},u_{\theta},u_{\phi},b)$ is a vector of data
variables at each grid point, of size
$4N$, where $N=N_{x}N_{y}N_{z}$ is the total number of grid points. $M$ is the number of
spherical harmonics for which we compute the wave amplitudes, which is
usually taken to be one. $\mathbf{x}=(A_{in},A_{out})$ is a vector of size $4M < 4N$, whose components are the
(complex) IW/OW amplitudes (for each $l$ and $m$ value), whose values
are to be determined. The matrix $A$ contains the IW and OW radial
functions and spherical harmonic functions, evaluated at the selected
grid points using the linear solution in \S~\ref{lineartheory3D}, and has size $4N\times 4M$.

We use a  method of least squares to fit our model to the data \citep{Press1992}. 
This finds the best fit between the linear model data and
simulation data, and computes the solution for which the sum of the squared
residuals is its least value. 
This problem has a unique solution if the $4N$ columns of
the matrix $A$ are linearly independent, which is true in our case,
since $W[J_{\nu}(kr)+iY_{\nu}(kr),J_{\nu}(kr)-iY_{\nu}(kr)]=\frac{4}{i\pi r}
\ne 0$, $\forall r$. The solution is found by solving the normal equations 
\begin{eqnarray}
\hat{\mathbf{x}} = (A^{T} A)^{-1} A^{T} \mathbf{y}.
\end{eqnarray}
We take into account radial variations in the amplitudes by splitting
up the region inside the forcing region into a set of concentric spherical
shells of thickness $\delta r \sim \lambda_{r}/2$, after removing an inner
region of a few grid cells. This approach assumes the solution is locally
independent of $r$, hence we can ignore radial derivatives of
the amplitudes within each shell. This is not valid in regions where
the solution varies rapidly. In addition, we speed up computation by
stepping over the grid points in each Cartesian direction, by factors
$i_{step},j_{step},k_{step}$, chosen to take values between 1 and
10. We always ensure that sufficient grid points are available in each shell to
accurately compute the amplitudes.

A Matlab routine which reads in SNOOPY data and solves the linear least
squares problem has been written, and tested with various
combinations of analytic IW/OW solutions. We compute the reflection
coefficient $\mathcal{R}=A_{out}/A_{in}$. For perfect standing waves, $A_{in} =
A_{out}$, and $\mathcal{R}=1$. If the IW is entirely
absorbed at the centre, then $\mathcal{R}=0$.

The main disadvantage of our approach is that $\omega\ne 1, l\ne2, m\ne2$ components also contribute to the
amplitudes. This problem can be ignored if we trust the computed values of
$\mathcal{R}$ only where the solution is well described by the
linear solution, i.e., far from the wave breaking and forcing
regions. We therefore reconstruct $u_{r}$, and plot it along the $x$-axis,
$u_{\theta}$ along the line $y=z$, $u_{\phi}$ along the line $y=x$,
and $b$ along the line $y=x$. All components of the solution are not
zero for all radii along these lines when $\xi = 0$ (this requires
shifting the phase of the linear model solution when the wave phase is
nonzero in the simulation data). The simulation variables and
  IW/OW solutions are plotted
  using this method in Figs.~\ref{256lam15fr01t49AinAout} and \ref{256lam15fr1t45AinAout}.
Comparing the reconstructed
solution with the simulation data allows us to check whether the
deconstruction has worked.

\section{Analytic calculation of $F$ in the launching region}
\label{energyfluxcalculation}

In this section, we adopt the notation of OL04, to avoid reproducing the
results in that paper, since we are simply extending the results
of their \S 4.4. In the launching region, we want to match the solutions of the
RZ and CZ. Near the boundary $r=r_{b}$, 
we assume $N^{2} \propto
(r_{b}-r)^{\alpha}$, where $\alpha \in \mathbb{R}^{+}$, for the moment
left unspecified. The characteristic radial extent of the transition
region is of order
$\epsilon^{\beta}$, where $\beta \in \mathbb{R}^{+}$, and $\epsilon$
is the ratio of the spin frequency to the dynamical frequency, which
is much smaller than unity for slowly rotating bodies. We write
\begin{eqnarray}
r = r_{b} - \epsilon^{\beta} x,
\end{eqnarray}
where $x$ is an inner variable, which is of order unity in the
launching region. We can write
\begin{eqnarray}
N^{2} = \epsilon^{\alpha \beta} \mathcal{D} x^{\alpha} +
O(\epsilon^{2\alpha \beta}),
\end{eqnarray}
where $\mathcal{D} = \frac{d N^{2}}{d x^{\alpha}} > 0$. 
If we pose the perturbation expansions
\begin{eqnarray}
u_{r}^{\prime} &\sim& \epsilon^{3-\beta(1+\alpha)} \bar{u}_{r}^{\prime}(x,\theta), \\
u_{\theta}^{\prime} &\sim& \epsilon \bar{u}_{\theta}^{\prime}(x,\theta), \\
u_{\phi}^{\prime} &\sim& \epsilon\bar{u}_{\phi}^{\prime}(x,\theta), \\
\rho^{\prime} &\sim& \epsilon^{2-\beta}\bar{\rho}^{\prime}(x,\theta), \\
p^{\prime} &\sim& \epsilon^{2}\bar{p}^{\prime}(x,\theta), \\
\Phi^{\prime} &\sim& \epsilon^{2} \check{\Phi}^{\prime}(r_{b},\theta) + \epsilon^{2+\beta} \bar{\Phi}^{\prime}(x,\theta),
\end{eqnarray}
then we obtain the linearised system of equations 111-116 from OL04,
except that we replace $x$ by $x^{\alpha}$ in the buoyancy equation (Eq.~115).
A linearised equation for the modified pressure ($\bar{W} =
\frac{\bar{p}^{\prime}}{\rho_{0}}-\frac{p_{(e)}^{\prime}\rho_{2}}{\rho_{0}^{2}}
+\check{\Phi}^{\prime}$) in the
launching region can now be derived, for which 
\begin{eqnarray}
\mathcal{L} \bar{W} + \frac{r^{2}}{\mathcal{D}}\partial_{x} \left(x^{-\alpha}
\partial_{x} \bar{W}\right) = 0,
\end{eqnarray}
where all coefficients are evaluated at $r=r_{b}$. This equation can be
solved using the separation of variables
\begin{eqnarray}
\bar{W} = \sum_{i} f^{(i)}(z)w_{i}(\theta ).
\end{eqnarray}
The operator $\mathcal{L}$ contains only angular derivatives, and
$\mathcal{L}\bar{w}_{i} = \lambda_{i}\bar{w}_{i}$, where $\lambda_{i} =
\frac{k^{2}r^{2}}{N^{2}}$, and $k$ is the WKB wavenumber of the waves. We thus obtain the equation
\begin{eqnarray}
f^{(i)}(z) + \partial_{z} \left(z^{-\alpha}
\partial_{z} f^{(i)}(z)\right) = 0,
\end{eqnarray}
where we have defined $z=\kappa_{i} x$, and the length scale
\begin{eqnarray}
\kappa_{i}^{-1} = \left(\frac{\mathcal{D}\lambda_{i}}{r^{2}}\right)^{-\frac{1}{2+\alpha}}.
\end{eqnarray}
In the text this lengthscale is referred to as $L_{launch}$. The solution of this equation can be written down as a linear
combination of Bessel functions of the first and second kinds, of
order $\frac{\alpha+1}{\alpha+2}$:
\begin{eqnarray}
\nonumber
  f^{(i)}(z) &=& a_{i} z^{\frac{\alpha+1}{2}} \left\{ J_{\frac{\alpha+1}{\alpha+2}}\left(\frac{2}{2+\alpha}
      z^{\frac{2+\alpha}{2}}\right) + \right. \\ && \hspace{2cm} \left. s_{i} i Y_{\frac{\alpha+1}{\alpha+2}}\left(\frac{2}{2+\alpha}
      z^{\frac{2+\alpha}{2}}\right)  \right\},
\end{eqnarray}
so that the complete solution in the transition region is
\begin{eqnarray}
\nonumber
\bar{W} &=& \sum_{i}a_{i}
z^{\frac{\alpha+1}{2}} \left\{J_{\frac{\alpha+1}{\alpha+2}}\left(\frac{2}{2+\alpha}
      z^{\frac{2+\alpha}{2}}\right) + \right. \\  &&\left.
\hspace{1.9cm} s_{i} i Y_{\frac{\alpha+1}{\alpha+2}}\left(\frac{2}{2+\alpha}
      z^{\frac{2+\alpha}{2}}\right)  \right\}w_{i}(\theta)
\end{eqnarray}
This solution should match onto the WKB solution in the RZ
when $x \gg 1$, constraining $\beta$ as a function of $\alpha$. This
arises from considering the asymptotic form of the phase at large $z$,
whose radial derivative should match the WKB wavenumber in the
RZ. From this we find $\beta = 2/(2+\alpha)$.

If we from now on restrict ourselves to $\alpha = 1/2$, then the
corresponding complete solution in the transition region is the sum
\begin{eqnarray}
\bar{W} = \sum_{i}a_{i} z^{\frac{3}{4}}
\left\{ J_{\frac{3}{5}}\left(\frac{4}{5}z^{\frac{5}{4}}\right) + s_{i} i J_{\frac{3}{5}}\left(\frac{4}{5}z^{\frac{5}{4}}\right)  \right\}w_{i}(\theta).
\end{eqnarray}
Similarly we can write the radial displacement as
\begin{eqnarray}
\label{besseltransitionsolution}
\bar{\xi}_{r}^{\prime} = \sum_{i} b_{i}\sqrt{z}\left\{ J_{\frac{2}{5}}\left(\frac{4}{5}z^{\frac{5}{4}}\right)
  \pm i t_{i} J_{\frac{2}{5}}\left(\frac{4}{5}z^{\frac{5}{4}}\right)\right\}w_{i}(\theta).
\end{eqnarray}
for a superposition of wave solutions in this model.

Deep in the RZ, our solution should reduce to a wave with an inwardly directed group
velocity, for which $s_{i} = \pm 1$ (sign depending on that of the frequency). The asymptotic forms of the
radial part of this solution at large $z$ is
\begin{eqnarray}
\nonumber
&& z^{\frac{3}{4}}\left\{ J_{\frac{3}{5}}\left(\frac{4}{5}z^{\frac{5}{4}}\right)
  \pm i J_{\frac{3}{5}}\left(\frac{4}{5}z^{\frac{5}{4}}\right)
\right\} \sim \\ && \hspace{2.5cm}
(-1)^{\frac{9}{20}}\sqrt{\frac{5}{2 \pi}} z^{\frac{1}{8}} \exp \left\{
  \mp i \frac{4}{5}z^{\frac{5}{4}}\right\}.
\end{eqnarray}
Matching of $W$ at $r=r_{b}$ requires the solution in the CZ
to be continuous with the solution in the transition region at $z=0$. This
determinines the amplitudes $a_{i}$. The inward energy flux in (inertia-)gravity
waves is found from evaluating
\begin{eqnarray}
F &=& \pi \int_{0}^{\pi}
\mathrm{Re}\left[\left(\frac{\omega}{\hat{\omega}}\right)p^{\prime
    *}u_{r}^{\prime}\right]r^{2}\sin \theta d\theta \\ \nonumber
&=& \epsilon^{\frac{19}{5}}
\frac{\left(\frac{2}{5}\right)^{\frac{1}{5}}\pi^{2}}{\left[\Gamma\left(\frac{3}{5}\right)\right]^{2}}
\rho_{0}
\omega_{(1)}\mathrm{sgn}(\hat{\omega})\left(\frac{r^{2}}{\mathcal{D}}\right)^{\frac{2}{5}}
\\ && \times\sum_{i}\lambda_{i}^{\frac{3}{5}} \frac{\vert\int_{0}^{\pi}w_{i}^{*}\left[\check{W}(r_{b},\theta)-\bar{W}^{(p)}\right]\sin
\theta d \theta\vert^{2}}{\int_{0}^{\pi}|w_{i}|^{2}\sin\theta
d\theta},
\end{eqnarray}
which can be compared with OL04 Eq.~124. We find in general that
$F\propto \omega^{\frac{8+3\alpha}{2+\alpha}}$. Thus we see that the
main change in the energy flux from modifying the profile of $N^{2}$
in the transition region is to change its frequency dependence. There
are also $O(1)$ changes to the numerical factors.

%\newcommand{\bibfont}{\small}
%\setlength{\bibsep}{0pt}
%\bibliography{tidbib}
%\bibliographystyle{mn2e}

\end{document}